\documentclass[draftcls,twoside,preprint, onecolumn,letter,11pt]{IEEEtran}

\usepackage{cite}

\ifCLASSINFOpdf
\else
\fi

\usepackage[cmex10]{amsmath}

\ifCLASSOPTIONcompsoc
\usepackage[caption=false,font=normalsize,labelfont=sf,textfont=sf]{subfig}
\else
\usepackage[caption=false,font=footnotesize]{subfig}
\fi

\usepackage{graphicx}
\usepackage{epstopdf}
\usepackage{myMacros}
\usepackage{amsthm}
\usepackage{multirow}

\newcommand{\Fig}[1]{Fig.~\ref{#1}\xspace}

\newcommand{\Xsq}{\ensuremath{X_{e}^2}}
\newcommand{\X}{\ensuremath{X_{e}}}

\newcommand{\Ex}[1]{\mathop{\bf E\/} \left[ #1 \right]}
\newcommand{\ExP}[1]{\mathop{\bf E'\/} \left[ #1 \right]}

\newcommand{\La}[1]{\ensuremath{#1^{*}(s)}}
\newcommand{\SumNX}{\ensuremath{\sum\nolimits_{i=1}^{N(X)}}}

\newcommand{\muOFF}{\ensuremath{ \mu_{\msf{off}}}}
\newcommand{\muON}{\ensuremath{ \mu_{\msf{on}}}}

\newcommand{\theTilO}{\ensuremath{\tilde{\Grq}_1}}

\newcommand{\theSo}{\ensuremath{\Grq_1^{\msf{S}}}}

\newcommand{\cSo}{\ensuremath{c_1^{\msf{S}}}}
\newcommand{\cSe}{\ensuremath{c_2^{\msf{S}}}}

\newcommand{\cRo}{\ensuremath{c_1^{\msf{R}}}}
\newcommand{\cRe}{\ensuremath{c_2^{\msf{R}}}}

\newcommand{\Fthe}{\ensuremath{F_{\GrQ}}}
\newcommand{\FtheP}{\ensuremath{F'_{\GrQ}}}
\newcommand{\fthe}{\ensuremath{f_{\GrQ}}}
\newcommand{\theMax}{\ensuremath{\Grq_{\msf{up}}}\xspace}
\newcommand{\xLo}{\ensuremath{x_{\msf{lo}}}\xspace}
\newcommand{\xUp}{\ensuremath{x_{\msf{up}}}\xspace}
\newcommand{\cLo}{\ensuremath{c_{\msf{lo}}}\xspace}
\newcommand{\cUp}{\ensuremath{c_{\msf{up}}}\xspace}
\newcommand{\ON}{\ensuremath{\msf{ON}}\xspace}
\newcommand{\OFF}{\ensuremath{\msf{OFF}}\xspace}
\newcommand{\Exp}{\ensuremath{\msf{Exp}}\xspace}
\newcommand{\Erl}{\ensuremath{\msf{Erl}}\xspace}
\newcommand{\UniExp}{\ensuremath{\msf{UniExp}}\xspace}
\newcommand{\ErlExp}{\ensuremath{\msf{ErlExp}}\xspace}
\newcommand{\ExpErl}{\ensuremath{\msf{ExpErl}}\xspace}
\newcommand{\Oo}{\ensuremath{\mathcal{O}_1}\xspace}
\newcommand{\Oe}{\ensuremath{\mathcal{O}_2}\xspace}

\newcommand{\EX}{\mathop{\bf E\/} \left[ \X \right]}
\newcommand{\ET}{\mathop{\bf E\/} \left[ T\left( p_1^*\right) \right]}
\newcommand{\ETn}{\mathop{\bf E\/} \left[ T\left( p_1\right) \right]}
\newcommand{\ETt}{\mathop{\bf E\/} \left[ T\left( p_1^{t} \right) \right]}

\newcommand{\GamN}{\ensuremath{\frac{c_2 - c_1}{\Ex{T\left( p_1 \right)} - \Ex{X} }}} 
\newcommand{\theBarLo}{\ensuremath{\bar{\Grq}( p^{\Gre} )\xspace}}

\begin{document}
%
\title{Optimal Pricing Effect on Equilibrium Behaviors of Delay-Sensitive Users in Cognitive Radio Networks }


\author{Nguyen~H.~Tran,~\IEEEmembership{Member,~IEEE,}
        Choong Seon Hong,~\IEEEmembership{Senior Member,~IEEE,}
        Zhu Han,~\IEEEmembership{Senior Member,~IEEE,}
        and~Sungwon~Lee,~\IEEEmembership{Member,~IEEE}
 \thanks{N.~H.~Tran, C.~S.~Hong and S.~Lee  are with the Department
 of Computer Engineering, Kyung Hee University, Korea (email: \{nguyenth, cshong, drsungwon\}@khu.ac.kr). 
 
 Z. Han is with the Electrical and Computer Engineering Department,
University of Houston, Houston, USA (email: zhan2@uh.edu).
 }}
 


\maketitle

\begin{abstract}
This paper studies price-based spectrum access control in cognitive radio networks, which characterizes network operators' service provisions to delay-sensitive secondary users (SUs) via pricing strategies. Based on the two paradigms of shared-use and exclusive-use dynamic spectrum access (DSA), we examine three network scenarios corresponding to three types of secondary markets.  In the first monopoly market with one operator using opportunistic shared-use DSA, we study the operator's pricing effect on the equilibrium behaviors of self-optimizing SUs in a queueing system. 
We provide a queueing delay analysis with the general distributions of the SU service time and PU traffic using the renewal theory. In terms of SUs, we show that there exists a unique Nash equilibrium in a non-cooperative game where SUs are players employing individual optimal strategies. We also provide a sufficient condition and iterative algorithms for equilibrium convergence. In terms of operators, two pricing mechanisms are proposed with different goals: revenue maximization and social welfare maximization. In the second monopoly market, an operator exploiting exclusive-use DSA has many channels that will be allocated separately to each entering SU. We also analyze the pricing effect on the equilibrium behaviors of the SUs and the revenue-optimal and socially-optimal pricing strategies of the operator in this market. In the third duopoly market, we study a price competition between two operators employing shared-use and exclusive-use DSA, respectively, as a two-stage Stackelberg game. Using a backward induction method, we show that there exists a unique equilibrium for this game and investigate the equilibrium convergence.
\end{abstract}


%
\IEEEpeerreviewmaketitle

\section{Introduction}
\begin{figure}[!t]
\centering
\includegraphics[width=0.67\textwidth]{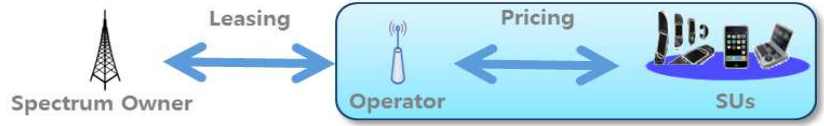}
\caption{ Three-tier dynamic spectrum market.}
\label{F:3layers}
\end{figure}

The concept of dynamic spectrum access (DSA) has been emerging as a new approach for efficient utilization of scarce wireless spectrum that is conventionally controlled via static licensing. DSA enables secondary (unlicensed) users (SUs) to flexibly access underutilized legacy spectrum bands that are used sporadically by primary (licensed) users (PUs) \cite{McHenry2005}. This idea of reusing the legacy spectrum is receiving great supports thanks to the rapid development of software-defined radio and cognitive radio (CR) technologies \cite{Mitola1999}, which can vary parameters such as frequency, power, and modulation schemes through software. Most DSA approaches \cite{Zhao2007, Buddhikot2007} are connected to a three-tier model of the dynamic spectrum market in \Fig{F:3layers}, which includes three network entities: spectrum owners, secondary network operators\footnote{For brevity, henceforth we will use only ``operators''.} and CR-enabled SUs. The spectrum owner can temporarily lease their spectrum to the operators through periodic auctions. In each period, the winning operators can provide secondary services for SUs via their leased spectrum by charging an admission price to SUs for an economic return. 

Among the various DSA approaches, opportunistic shared-use (i.e. spectrum overlay) and dynamic exclusive-use models have been widely considered \cite{Zhao2007,Buddhikot2007, Hossain2009}. Opportunistic shared-use allows spectrum owners employ  ``interruptible leasing'' to operators, a model that forces operators to provide secondary services without harming the operations of PUs on the leased spectrum. Dynamic exclusive-use, on the other hand, allows spectrum owners to dynamically transfer  spectrum-usage rights to operators, a model that enables operators to lease parts of temporarily unused spectrum (i.e. no PUs operations) from spectrum owners for secondary services provision.

In this paper, we study  pricing-based spectrum access control of SUs by operators, which is shown in the shaded region of \Fig{F:3layers}. We consider two kinds of operators that employ the two different DSA approaches of opportunistic shared-use and dynamic exclusive-use\footnote{For brevity, henceforth we will use only ``shared-use'' and ``exclusive-use''.}, respectively. Their service provision to SUs are controlled through pricing-based methods and are considered in three secondary-market types: shared-use monopoly, exclusive-use monopoly, and duopoly pricing competition.
%
%

In the first monopoly market, by considering delay-sensitive SUs that share a PU's single channel controlled by a shared-use operator, we examine the effect of the operator's pricing on the equilibrium behaviors of non-cooperative SUs. This behavior is represented by a SU's spectrum access decision upon arrival, which entails either joining the list of other SUs that also want to share the same channel or balking. In terms of SUs, we first introduce an individual optimal strategy employed by each SU as a player in a non-cooperative game in order to make its spectrum access decision, based on a utility function that captures the delay-sensitivity heterogeneity of SUs. We next show that there exists a unique Nash equilibrium in this game. In order to employ the individual optimal strategy, each SU has to evaluate its mean queueing delay of a virtual queue that is used to model a congestion effect that occurs when many SUs intend to share the same PUs' channel. By using the renewal theory, we provide a queueing delay analysis based only on  statistical information of the PUs' activities and SUs' service time. Finally, we examine equilibrium dynamics through iterative algorithms and provide a sufficient condition for equilibrium convergence. In terms of the operators, we devise two pricing mechanisms. For the case of the operator being a commercial planner, we propose a revenue-optimal pricing policy to maximize the operator's revenue by solving a convex optimization problem. For the case of the operator being a social planner, we propose a socially-optimal pricing policy to maximize network social welfare by solving an equivalent convex problem.

In the second monopoly market, each entering delay-sensitive SU is allocated a dedicated channel from a list of channels managed by an exclusive-use operator. Similarly to the first monopoly market, we also study the effect of the operator's pricing  on the equilibrium behaviors of the non-cooperative SUs.  Since queueing delays of the SUs' jobs are equal to their service time due to these dedicated channels, the analytic results of the SU's equilibrium behaviors, revenue-optimal pricing and socially-optimal pricing can also be obtained in this market. 

In the third duopoly market, we treat the competition between shared-use and exclusive-use operators as a two-stage Stackelberg game, where two operators aim to maximize their revenues through pricing and each SU will make its decision of which operators to join based on the prices charged by the two operators. We first show that there exists a unique equilibrium for this game. We then explore the equilibrium dynamics via the iterative algorithms and provide a sufficient condition for the equilibrium convergence.

The rest of this paper is organized as follows. Related work is reviewed in Section~\ref{S:relWork}, and a system model is described in Section~\ref{S:sysModel}. We analyze the shared-use monopoly in Section~\ref{S:SP1}, the exclusive-use monopoly in Section~\ref{S:SP2}, and the duopoly market in Section~\ref{S:Duo}. Numerical studies are provided in Section~\ref{S:numResult}. Finally, we conclude our work in Section~\ref{S:conclusion}.

\section{Related Work} \label{S:relWork}
There is an increasing interest in the three-tier model of the dynamic spectrum market, which  consists of interactions among spectrum owners, operators, and SUs \cite{Duan2011, Kim2011a, Duan2012, Jia2008a, Niyato2009}. However, the model of price-based spectrum access control between  operators and multiple SUs has attracted the  interests. The studies in \cite{Xing2007, Niyato2008, Yang2011} examined  cases in which multiple operators compete for SUs, whereas \cite{Ileri2005} considered multiple SUs competing to access operators' channels. The authors of \cite{Niyato2009a} considered short-term spectrum trading across multiple PUs and SUs. \cite{Zhang2009} considered interactions between an  operator and multiple SUs.  

Nevertheless, most of these works focused on  operators' pricing and the responses of SUs via their demand functions (e.g. the SUs' required bandwidth). In this paper, we focus on the pricing mechanisms and their impact on the equilibrium behaviors of SUs in a queueing system, which can be traced back from the original work of \cite{Naor1969, Edelson1975} and surveyed in the monograph \cite{Hassin2003}. To that end, there are many works which attempt to balance  between observable and unobservable queueing models. The observable queue system \cite{Li2011,Economou2008, Do2012}  requires either a centralized control server or a feedback mechanism with time overhead. In contrast, we use the unobservable queue system, which appropriately models the non-cooperative and distributed nature of SUs where SUs have no information about each other. Recent work on this paradigm of CR applications used server vacations or breakdowns in their queueing systems to model the opportunistic shared-use approach \cite{Jagannathan2011, Li2011}. In \cite{Li2011}, a queue was centrally controlled so that the current queue length could be observed for the SU decision making process. This work also used a discrete-time model where all distributions of arrivals and services were simply limited to a Bernoulli distribution in order to facilitate  analysis. The work in \cite{Jagannathan2011} used the unobservable and continuous-time models; however, the inter-arrival times and services were restricted to the exponential distributions for  ease of analysis. Both papers also assumed homogeneous SUs and the Markovian channels to model the PU traffic. Our work not only considers the heterogeneous SUs in terms of delay sensitivity, but also provides a queueing model where the PUs' channels and the service distributions of the SUs can be general.

\section{System Model} \label{S:sysModel}
We assume that there are two wireless network operators providing different DSA models. The first operator, denoted by \Oo, uses the shared-use model, whereas the second operator, denoted by \Oe, employs the exclusive-use model. We consider a network that consists of either one of the operators (i.e. monopoly) or both of them (i.e. duopoly), which corresponds to three types of secondary markets (cf. \Fig{F:model}). A stream of SUs is assumed to arrive at the network and each of them will make a decision as to whether or not to join an operator (in the case of a monopoly) or which operator to join (in the case of a duopoly).

\subsection{SUs}
We proceed to describe  important parameters of the SUs.
\subsubsection{Arrival Rates and Service Time} We assume that the SUs arrive at the network according to a Poisson process with rate \Grl. Each SU is associated with a distinct job (e.g. a packet, session, or connection) that it carries upon arrival. The service time to complete a job is represented by a random variable $X$ with a  probability density function (pdf) $f_X(x)$. This service time is assumed to be independent of the arrival process. 

\subsubsection{Delay-Sensitive User Types} Since the SUs are assumed to carry delay-sensitive traffic, each job is attached to a specific application type characterized by a parameter \Grq. This parameter represents an individual preference that reflects the delay sensitivity of the SU's application. The value of \Grq varies across job types, capturing \emph{SUs' heterogeneity}. Individual values of \Grq are private, but their cumulative distribution function, denoted by $F_{\GrQ}(.)$, is known. We also assume that this parameter follows a uniform distribution on $ \/[0,\theMax\/] $, which is common in the literature \cite{Manshaei2008, Chau2010, Ren2011}.  The relationship between \Grq and application types is presented through some examples: many multimedia applications with stringent delay requirements will have high values of \Grq; on the other hand, applications with $\Grq$ equal to zero are insensitive to delay.

\subsubsection{Individual Utilities} The value \Grq of a SU is realized at the instant it arrives (not before). This so-called type-\Grq SU then must make a decision: either join the network or balk. The utility of any balking SU is set to zero. For a type-\Grq SU that joins operator $\mathcal{O}_i$, its utility function is given by
\begin{align}
	U_i(\Grq) = V - \Grq d_i - c_i,	\quad i = 1, 2.	\label{E:Ui}
\end{align}
This utility function, which is widely used in the literature \cite{Chau2010, Gibbens2000}, captures the balance between a reward $V$ and a total cost $\Grq d_i + c_i$ that a SU undertakes once it decides to join the system. The reward $V$, which is assumed to be independent of SUs' application type, represents a benefit of a SU for accessing the service \cite{Gibbens2000}. The total cost consists of two elements: the admission price $c_i$ charged by $\mathcal{O}_i$ (i.e. the SUs are price-takers), and the waiting cost $\Grq d_i$ of a job that spends a delay $d_i$. In this waiting cost, the parameter $\Grq$ can be interpreted as a waiting cost per unit time, an interpretation that adheres to the delay-sensitivity mindset of \Grq: a higher waiting cost per unit time induces more negative effects of the delay, which shows more sensitivity to delay. We also assume that the unit of $\Grq$ is chosen such that $\Grq d_i$ has the same unit of $V - c_i$.

\subsection{Shared-Use Operator (\Oo)}
The operator \Oo is assumed to own a single channel. This channel is licensed to  legacy PUs, and is shared opportunistically by multiple SUs based on an admission price charged by \Oo. 
\begin{figure}[!t]
\centering
	\subfloat[]{\includegraphics[width=0.27\textwidth]{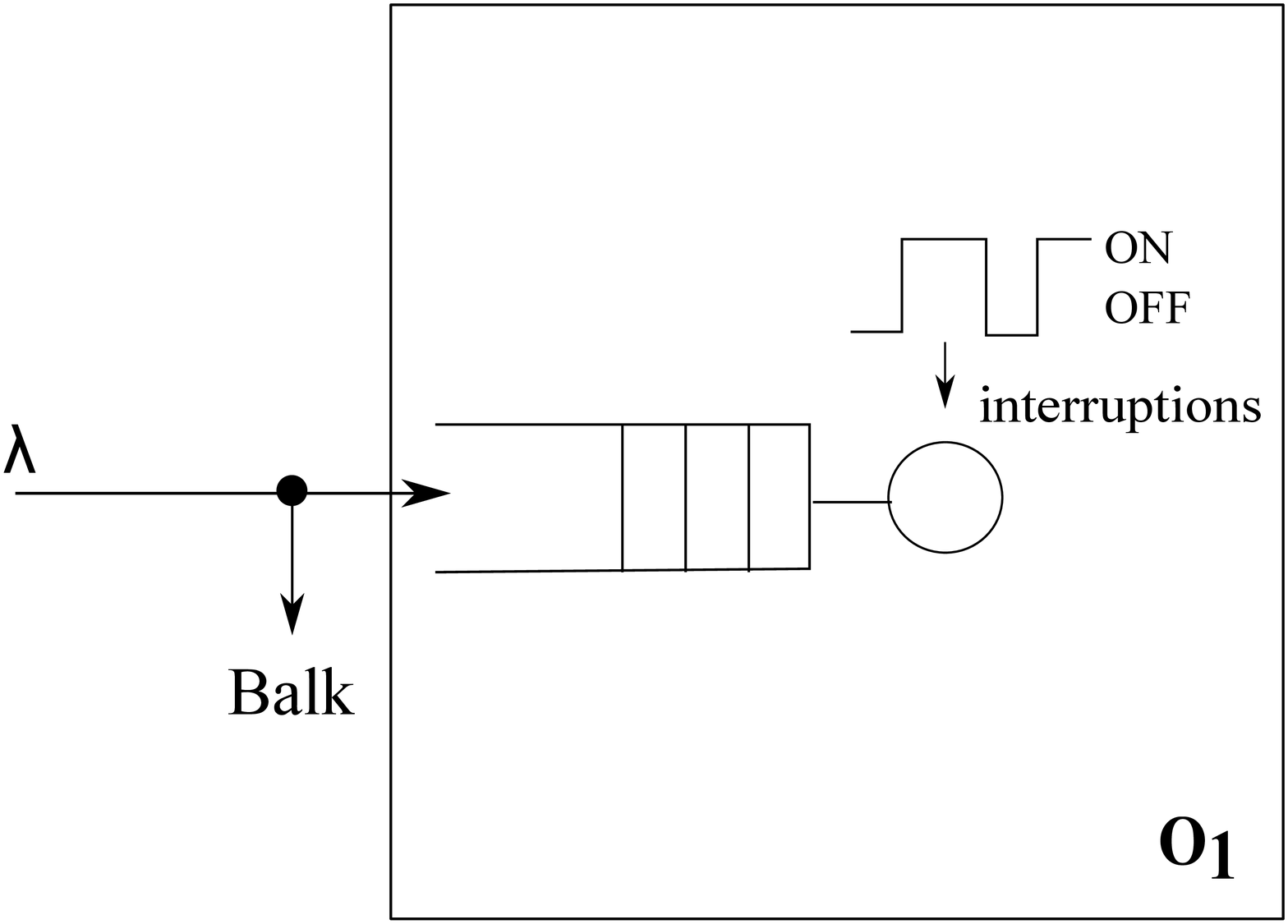}
	\label{F:O1}}
	\hfil
	\subfloat[]{\includegraphics[width=0.18\textwidth]{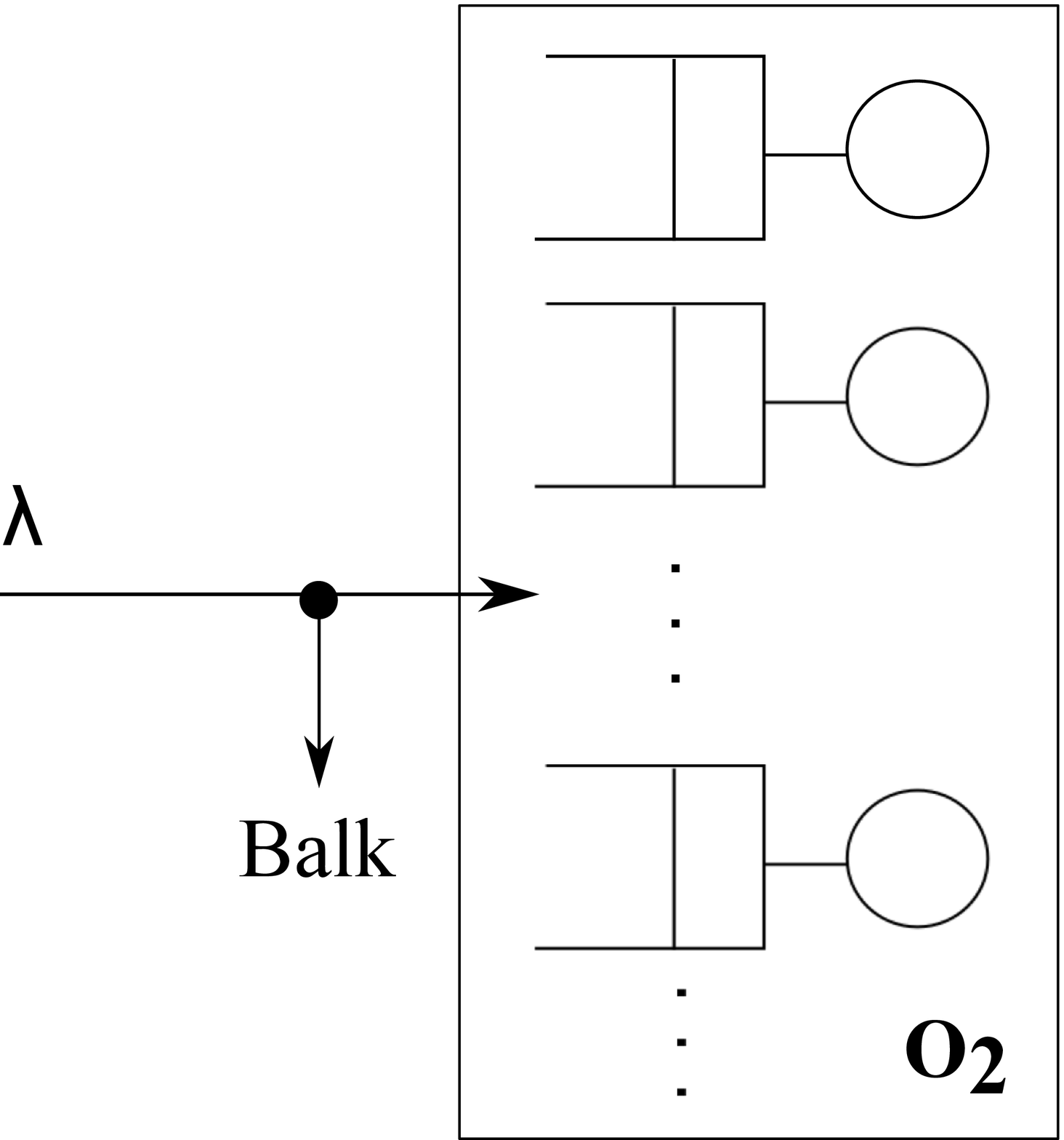}
	\label{F:O2}}
	\hfil
	\subfloat[]{\includegraphics[width=0.19\textwidth]{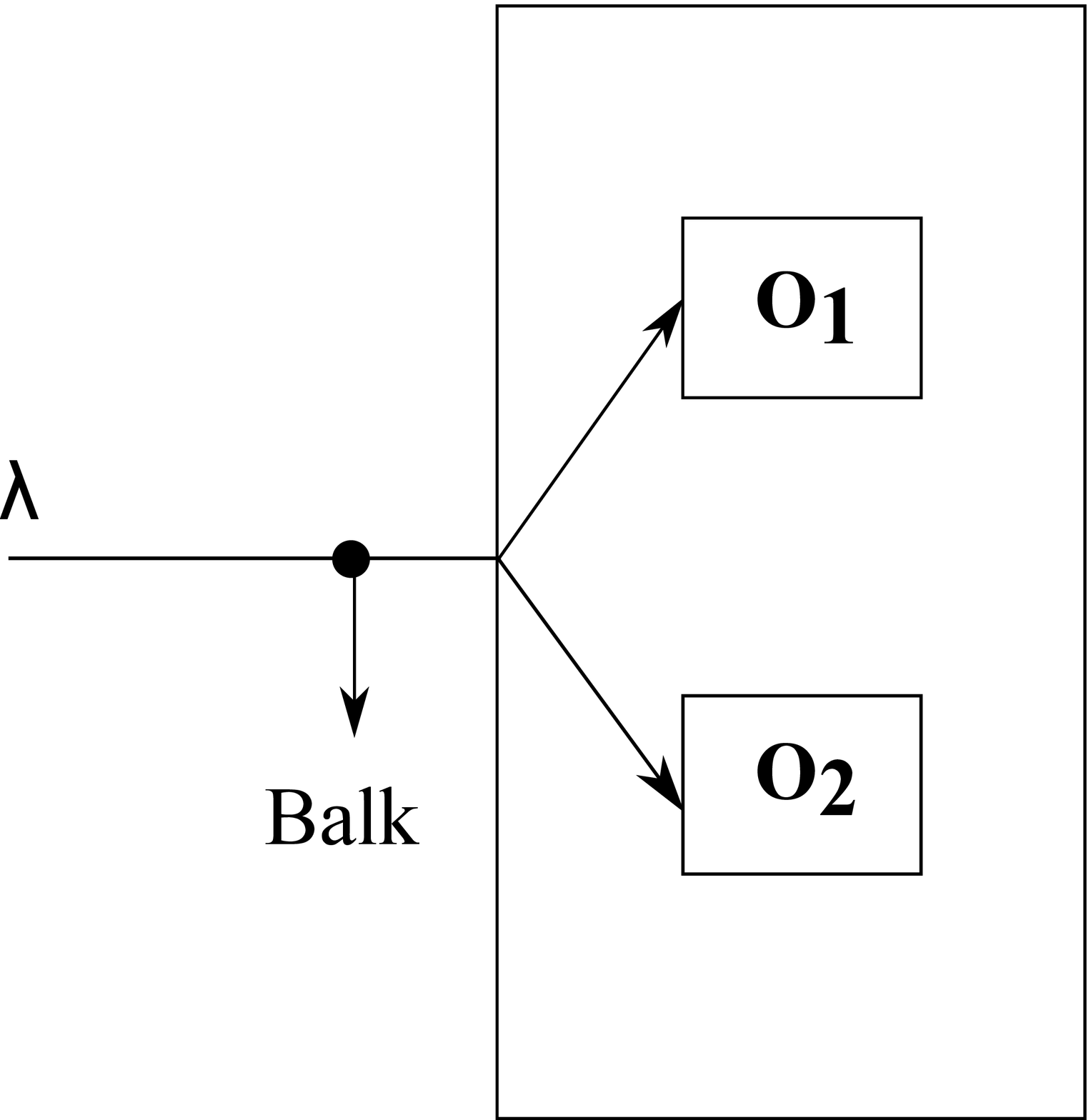}
	\label{F:duopoly}}
	\caption{Three secondary-market scenarios: (a) Shared-use Monopoly, (b) Exclusive-use Monopoly and (c) Duopoly.}
	\label{F:model}
\end{figure}
\subsubsection{PUs}
Traffic patterns of PUs on the licensed band can be modelled as an \ON-\OFF renewal process alternating between \ON (busy) and \OFF (idle) periods. We model the sojourn times of the \ON and \OFF periods as i.i.d. random variables $Y$ and $Z$, with the pdf $f_Y(y)$ and $f_Z(z)$, respectively. We assume that the \ON and \OFF periods are independent with SUs' arrival process and service time.

This \ON-\OFF process can be considered a channel model for the SU services.  This model captures the idle time period in which the SUs can utilize the channel without causing harmful interference to the PUs. We note that this PU traffic model is more general than other Markov \ON-\OFF models in which both busy and idle periods are restricted to the exponential distributions \cite{Jagannathan2011, Li2011,Geirhofer2008}.

\subsubsection{Steady-State Virtual Queue} Since many SUs may attempt to share the same licensed channel, congestion can occur, which will clearly affect the delay of each SU job.  Therefore, when a SU job arrives, it will evaluate its job's delay in a queue containing other SU jobs that also wish to use that licensed channel. This queue is only a \emph{virtual queue} because each SU cannot observe how many other jobs are waiting before its job (since SUs cannot know how many other SUs are trying to share the licensed channel).  Therefore, each SU forms a virtual queue based on the statistical information of $\Grl$, $f_X(x)$, $f_Y(y)$ and $f_Z(z)$, which are assumed to be estimated by existing methods \cite{Li2008}, to assess the mean queueing delay that its newly arrived job incurs. There are also many proposals in the CR literature which use the concept of virtual queue  to model the congestion effect \cite{Jagannathan2011, Shiang2008} in different contexts. Henceforth, we simply use ``queue'' to refer to this virtual queue. We can consider \Oo to be an M/G/1 queueing system (cf. \Fig{F:O1}) whose service time has a general distribution dictated by $f_X(x)$, $f_Y(y)$ and $f_Z(z)$ since a SU service occuring in \OFF periods can be interrupted by the returning of PUs in \ON periods. We denote $\Ex{T(\Grl)}$ the mean steady-state queueing delay (i.e. waiting time + service time) induced by an arrival rate $\Grl$. From \eqref{E:Ui}, the utility of a type-\Grq SU with \Oo is 
\begin{align}
	U_1(\Grq) = V - \Grq \Ex{T(\Grl)} - c_1.	\label{E:U1}
\end{align}

\subsection{Exclusive-Use Operator (\Oe)}
The operator \Oe is assumed to obtain (i.e. via leasing) the part of spectrum which is temporarily unused  by the spectrum owner. This spectrum chunk is divided into multiple bands that have the same bandwidth as the single band of \Oo. Since there is no PU traffic on these bands, SU services are not interrupted  in this case.

Whenever an arriving SU decides to join \Oe, the operator allocates a dedicated channel for the SU. We assume that \Oe always has enough dedicated bands to serve the SUs\footnote{This assumption can be relaxed by ``borrowing'' more channels from other homogeneous operators when \Oe lacks the dedicated channels \cite{Buddhikot2007}.}. Therefore, we can consider \Oe to be an M/G/$\infty$ queueing system (cf. \Fig{F:O2}) where  queueing delays of all SUs are equal to $\Ex{X}$. From \eqref{E:Ui}, the utility of a type-\Grq SU with \Oe is
\begin{align}
	U_2(\Grq) = V - \Grq \Ex{X} - c_2.	\label{E:U2}
\end{align}

\section{Type I: Shared-Use Monopoly Market} \label{S:SP1}
In this section, we first investigate the SUs' strategies with the mean queueing delay analysis, the Nash equilibrium, and the equilibrium convergence. We then examine the \Oo's optimal pricing policies in terms of the revenue and social maximization.
\subsection{SUs' Strategies}

\subsubsection{Nash Equilibrium}
We consider a stream of self-optimizing arriving SUs, which are concerned only with their own benefits. In the game theory context, the potential SUs behave like \emph{players} in a non-cooperative game, and the decisions regarding joining or balking are their \emph{strategies}. Specifically, upon arrival, each type-\Grq SU has to make a decision based on the joining probability $p(\Grq) \in \/[0, 1\/]$. Given the joining rule $\{p(\Grq), \, \Grq \geq 0\}$, the unconditional probability that a potential SU joins the monopoly \Oo is $\, p_1 = \int_{0}^{\infty} p(\Grq) d \Fthe (\Grq)$. With this joining rule, the actual arrival rate to the system is $\Grl p_1 $. Because \Grl is fixed,  we denote queueing delay by $\ETn$ rather than $\Ex{T(\Grl p_1)} $ for ease of presentation.

Since SUs are self-optimizing, each type-\Grq SU will choose its joining probability $p(\Grq)$ to maximize its expected utility $ p(\Grq) U_1(\Grq) + (1 - p(\Grq))0$, which corresponds to the following \emph{individual optimal strategy}.
\begin{definition}\label{D:IOS1}
An individually-optimizing type-\Grq SU that has $U_1(\Grq)  = V - \Grq \Ex{T(p_1)} - c_1$ will join \Oo
\begin{itemize}
	\item  with probability $p(\Grq)=1$ if $ U_1(\Grq) > 0 $, which requires
		\begin{align}
			 \Grq < \Grq_1(p_1), \text{ where } \Grq_1(p_1) \triangleq \frac{V - c_1}{\ETn},  \label{E:gam1}
		\end{align}
	\item with probability $p(\Grq)=0$, otherwise.
\end{itemize}
\end{definition}
Condition \eqref{E:gam1} states that when all SUs employ the individual optimal strategy, only a fraction of SUs that have \Grq values less than $\Grq_1(p_1)$ will join the \Oo. Since the unconditional joining probability $p_1$ can be considered to be the fraction of SUs that join \Oo, we have
\begin{align}
	p_1 &= \int_{0}^{\infty} p(\Grq) d \Fthe (\Grq)	=  \int_{0}^{\Grq_1(p_1)} d \Fthe (\Grq) = \Fthe \bigl( \Grq_1(p_1) \bigr).
\end{align}

Therefore, the equilibrium of the SUs' joining probability to \Oo is defined as follows.

\begin{definition} \label{D:EJP} 
$\, p_1^*$ is a Nash equilibrium of SUs' joining probability in a shared-use monopoly if it satisfies
\begin{align}
	p_1^* = \Fthe \bigl( \Grq_1(p_1^*) \bigr).	\label{E:fixpoint}
\end{align}
\end{definition}

This definition shows that once reaching an equilibrium, the fraction of joining SUs remains the same hereafter. The equilibrium $p_1^*$ is called a Nash equilibrium if at this point, no SU has any incentive to deviate from its strategy assuming that all other SUs continue to follow their strategies. The following theorem establishes the existence and uniqueness of the Nash equilibrium, of which the proof is provided in Appendix~\ref{A:T_Uniq}.
\begin{theorem} \label{T:Uniq}
For a given admission price $c_1$, there exists a unique Nash equilibrium of the SUs' joining probability $p_1^*$ in a shared-use monopoly market. 
\end{theorem}
\subsubsection{Queueing Delay Analysis}
\begin{figure}[!t]
\centering
\includegraphics[width=0.56\textwidth]{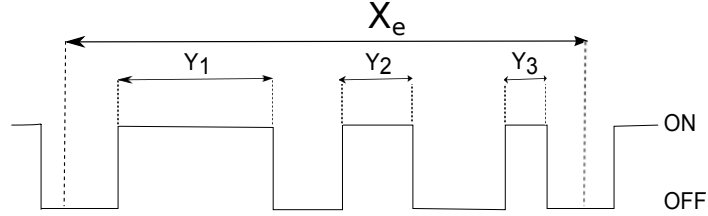}
\caption{ A sample \ON-\OFF process with a realization of an effective service time $X_e$ where its original SU service time is increased because of three interruptions from \ON periods $Y_1$, $Y_2$ and $Y_3$. }
\label{F:Xe}
\end{figure}
In order to perform its individual optimal strategy, each SU must estimate the mean queueing delay, which will be analyzed in the sequel.

We assume that a SU can use its spectrum sensing and handoff capabilities to detect and protect the PUs, respectively. Spectrum sensing is used to inform the SU whether the channel is busy or idle. Moreover, sensing errors are assumed to be negligible. When the channel is sensed to be idle, the SU job can be in service. When the channel is sensed to be busy, the spectrum handoff procedure is performed to return the channel to the PUs. This kind of the listen-before-talk channel access scheme has been adopted in the quiet period technique of the IEEE 802.22 standard \cite{Stevenson2009}. During the service time of a SU job, it is likely that the SU must perform multiple spectrum handoffs due to multiple interruptions from the returns of PUs, represented by \ON periods. Spectrum handoffs, which are employed to protect PU traffic and to provide reliable SU services, help SUs vacate the channel during \ON periods and resume their unfinished services after \ON periods end. Clearly, in the case of multiple spectrum handoffs, the original service time $X$ of the SU job is increased,  as illustrated in \Fig{F:Xe}, and this increased service time is called the \emph{effective service time} and is denoted by a random variable $\X$. 

We begin the analysis by denoting $\Ex{W(p_1)}$ the mean waiting time in the M/G/1 queueing system whose mean service time and arrival rate are $\Ex{\X}$ and  $\Grl p_1$, respectively. According to the Pollaczek-Khinchin formula \cite{Bertsekas1992}, the mean waiting time is 
\begin{align}
	\Ex{W(p_1)} = \frac{\Grl p_1 \Ex{\Xsq}}{2 ( 1 - \Grl p_1 \Ex{\X} )}.	\label{E:waiting}
\end{align}
Using the mean value analysis in \cite{Bertsekas1992}, we have the extended-value mean queueing delay as follows 
\begin{equation}  \label{E:qDelay}
   \Ex{T(p_1)} = \begin{cases}
               \Ex{W(p_1)} + \Ex{\X},   &\text{if } \Grl p_1 < 1/\Ex{\X};	\\
               \infty,	& \text{otherwise.}
           \end{cases}
\end{equation}
This extended-value queueing delay can eliminate the explicit condition $\Grl p_1 < 1/\Ex{\X}$ in our arguments hereafter.  The problem boils down to how to derive $\Ex{\X}$ and $\Ex{\Xsq}$, the first and second moments of the effective service time, respectively, in order to estimate the mean queueing delay. We proceed to use the renewal theory to derive these moments based on the statistical information of the SU service time and the \ON-\OFF process. 

\textbf{The First Moment of Effective Service Time.} Defining a random variable, $N(X)$, as the number of renewals (i.e. \ON periods) occurring in the interval $(0,\, X)$, we have
\begin{align}
\Ex{\X} &= \Ex{X + \SumNX Y_i}	= \Ex{X} + \Ex{ \SumNX Y_i }	= \Ex{X} + \Ex{Y} \Ex{ N(X) }, \label{E:Xe}
\end{align}
where the final equality occurs because $Y$ is independent of $X$. From \cite[pp. 45]{Cox1967}, we have
\begin{align}
	\Ex{N(X)\bigm\vert X = x } = \frac{x}{\Ex{Z}}.
\end{align}
As a consequence, $\Ex{ N(X) } = \frac{\Ex{X}}{\Ex{Z}}$, which is then substituted into \eqref{E:Xe} so as to obtain
\begin{align}
	\Ex{\X}	= \Ex{X} \left( 1 + \frac{\Ex{Y}}{\Ex{Z}}  \right).
\end{align}

\begin{figure*}[!t]
\centering
	\subfloat[$\muON = 0.5, \, \muOFF = 1.5$]{\includegraphics[width=0.48\textwidth]{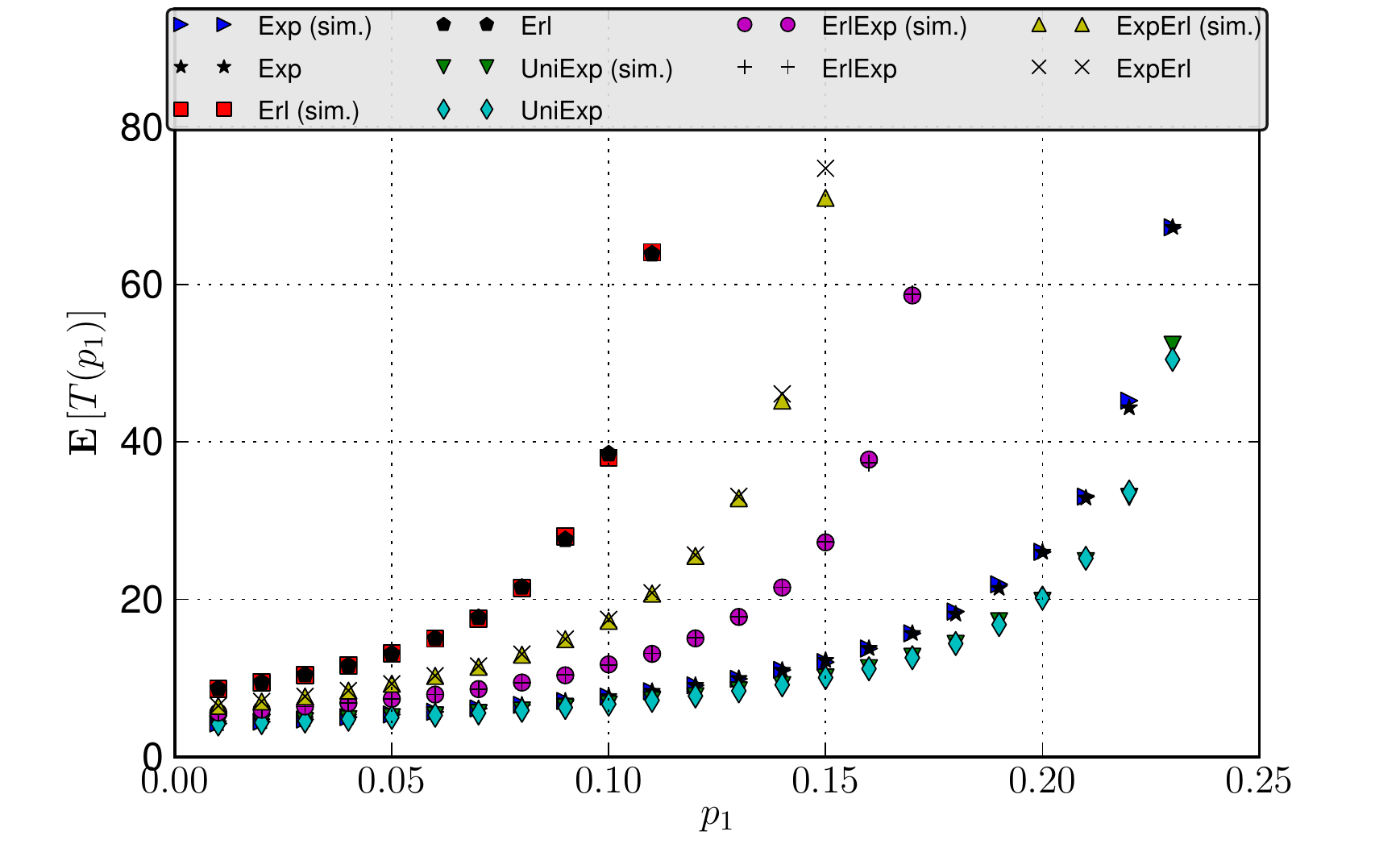}
	\label{F:simAnalQ0}}
	\hfil
	\subfloat[$\muON = 1.5, \, \muOFF = 0.5$]{\includegraphics[width=0.48\textwidth]{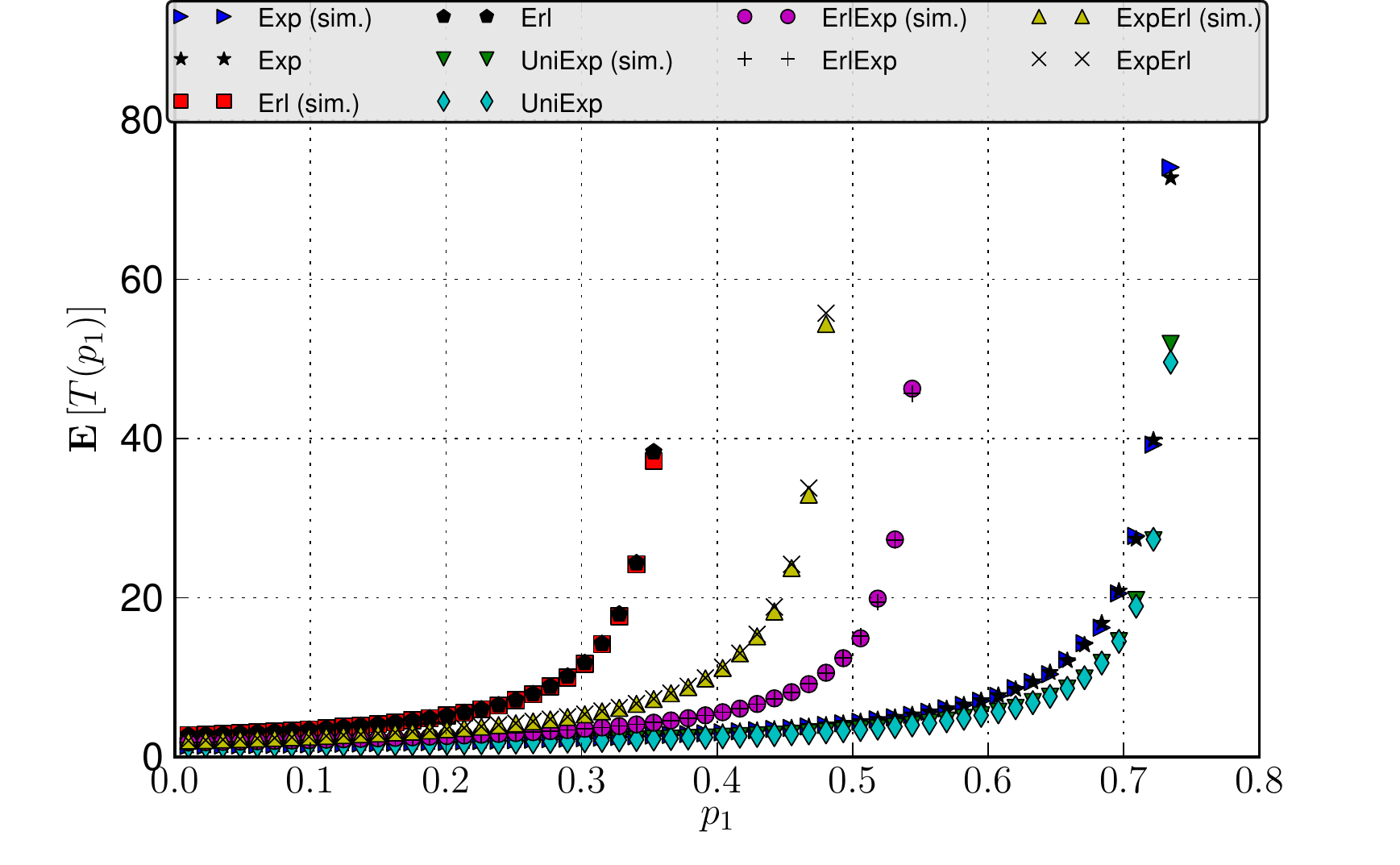}
	\label{F:simAnalQ1}}
	\caption{Mean queueing delay performance comparison: (a) heavy PU traffic model, (b) light PU traffic model.}
	\label{F:simAnalQ}
\end{figure*}

\textbf{The Second Moment of Effective Service Time}. We continue with
\begin{align}
	\Ex{\Xsq} &= \Ex{\left( X + \SumNX Y_i \right)^2}	= \Ex{X^2} + 2 \Ex{X \SumNX Y_i}	+ \Ex{\left( \SumNX Y_i \right)^2}.\label{E:XeSquare}
\end{align}	
Using the law of iterated expectations, the second term of the right side of \eqref{E:XeSquare} can be shown to be
\begin{align}
	\Ex{X \SumNX Y_i \mid X=x} &= x \Ex{N(x) Y}	= x \Ex{N(x)} \Ex{Y}	= x^2 \frac{ \Ex{Y} }{ \Ex{Z} }, 
\end{align}	
where we have the second equality because of the independence between $Y$ and $N(x)$. Hence, we obtain
\begin{align}	
		\Ex{X \SumNX Y_i} &= \Ex{X^2} \frac{\Ex{Y}}{\Ex{Z}}.	\label{E:term2}
\end{align}	
Next, we derive the third term on the right side of \eqref{E:XeSquare} as follows
\begin{align}	
		\Ex{\left( \SumNX Y_i \right)^2}	&= \Ex{ \SumNX Y_i^2 } + \Ex{ \sum\limits_{\{(i,j) \mid i \neq j \}} Y_i Y_j }	\nonumber\\
		&= \Ex{N(X)} \Ex{Y^2} + \Ex{Y}^2 \Ex{  \bigl(N(X) - 1\bigr) N(X) }.	\label{E:term3}
\end{align}	
We define $g(X \mid X=x) \triangleq \Ex{  \bigl(N(X) - 1\bigr) N(X) \bigm\vert X = x }$ and denote the Laplace transform of an arbitrary function $f(x)$ by \La{f}. Using the similar technique of deriving the variance of the number of renewals in \cite[pp. 55]{Cox1967}, we can easily obtain the following result
\begin{align}
	\La{g} = \frac{ 2 }{ s^2 \Ex{Z} } \frac{ \La{f_Z} }{  1 - \La{f_Z}  }.
\end{align}
An inverse Laplace transform can then be applied to $\La{g}$ so as to obtain $g(x)$. Therefore, $g(X)$ can be found correspondingly. From \eqref{E:XeSquare}, \eqref{E:term2}, and \eqref{E:term3}, we can see that $\Ex{\Xsq}$ is completely derived.

\textbf{Examples with Analysis and Simulation Comparisons.} \label{SSS:settings}
We supplement the queueing delay analysis through a performance comparison of the analysis and simulation by the following five examples.
\begin{enumerate}[(a)]
	\item $X, Y$ and $Z$ all have the exponential distributions with $f_X(x) = \mu_X e^{- \mu_X x}$, $f_Y(y) = \muON e^{- \muON y}$ and $f_Z(z) = \muOFF e^{- \muOFF z}$, respectively. This combination is termed \Exp, and we obtain
	\begin{align}
	\Ex{\X} &= \frac{1}{\mu_X}\left( 1 + \frac{\muOFF}{\muON} \right),	\\
	\Ex{\Xsq} &= \frac{2 \muOFF^2}{\muON^2 \mu_X^2}+\frac{2 \muOFF}{\muON^2 \mu_X}+\frac{4 \muOFF}{\muON \mu_X^2}+\frac{2}{\mu_X^2}.	
	\end{align}
	\item $X, Y$ and $Z$ all have the Erlang distributions with $f_X(x) = \mu_X^2 x e^{- \mu_X x}$, $f_Y(y) = \muON^2 y e^{- \muON y}$ and $f_Z(z) = \muOFF^2 z e^{- \muOFF z}$, respectively. This combination is termed \Erl, and we obtain
	\begin{align}
		\Ex{\X} &= \frac{2}{\mu_X}\left( 1 + \frac{\muOFF}{\muON} \right),	\\
		\Ex{\Xsq} &= \frac{6 \muOFF}{\muON^2 \mu_X}+\frac{12 \muOFF}{\muON \mu_X^2}+\frac{6}{\mu_X^2}+ \frac{8 \muOFF^3 (3 \muOFF+2 \mu_X)}{\muON^2 \mu_X^2 (2 \muOFF+\mu_X)^2}.
	\end{align}
	\item $X$ is uniformly distributed on $\/[\xLo, \xUp\/]$, whereas $Y$ and $Z$ have the exponential distributions with $f_Y(y) = \muON e^{- \muON y}$ and $f_Z(z) = \muOFF e^{- \muOFF z}$, respectively. This example is termed \UniExp, and we obtain
	\begin{align}
		\Ex{\X} &= \frac{\xLo + \xUp}{2}\left( 1 + \frac{\muOFF}{\muON} \right),	\\
		\Ex{\Xsq} &= \frac{N}{48 \muOFF \muON^2 (\xUp-\xLo)},
	\end{align}
where $N =  (\xUp^3 - \xLo^3 ) ( 4 \muOFF^3 +16 \muOFF^2 \muON + 16 \muOFF \muON^2) +  42 \muOFF^2(\xUp^2 - \xLo^2 ) + 6 \muOFF (\xUp - \xLo) +  3 (e^{-2 \muOFF \xLo} -  e^{-2 \muOFF \xUp}).$
	\item $X$ has the Erlang distribution with $f_X(x) = \mu_X^2 x e^{- \mu_X x}$, whereas $Y$ and $Z$ have the exponential distributions with $f_Y(y) = \muON e^{- \muON y}$ and $f_Z(z) = \muOFF e^{- \muOFF z}$, respectively. This combination is termed \ErlExp, and we obtain
	\begin{align}
		\Ex{\X} &= \frac{2}{\mu_X}\left( 1 + \frac{\muOFF}{\muON} \right),	\\
		\Ex{\Xsq} &= \frac{6 \muOFF^2}{\muON^2 \mu_X^2}+\frac{4 \muOFF}{\muON^2 \mu_X}+\frac{12 \muOFF}{\muON \mu_X^2}+\frac{6}{\mu_X^2}.	
	\end{align}
	\item $X$ has the exponential distribution with $f_X(x) = \mu_X e^{- \mu_X x}$, whereas  $Y$ and $Z$ have the Erlang distributions with $f_Y(y) = \muON^2 y e^{- \muON y}$ and $f_Z(z) = \muOFF^2 z e^{- \muOFF z}$, respectively. This combination is termed \ExpErl, and we obtain
		\begin{align}
		\Ex{\X} &= \frac{1}{\mu_X}\left( 1 + \frac{\muOFF}{\muON} \right),	\\
		\Ex{\Xsq} &= \frac{4 \muOFF^3}{\muON^2 \mu_X^2 (2 \muOFF+\mu_X)}+\frac{3 \muOFF}{\muON^2 \mu_X}+\frac{4 \muOFF}{\muON \mu_X^2}+\frac{2}{\mu_X^2}.	
		\end{align}
\end{enumerate}

In order to validate our queueing analysis, we simulate a single-server queue with service interruptions for the performance comparison. In all five examples, we fix $\Grl = 1$ and vary $p_1$ to adjust the traffic load into the queue. We set $\mu_X = 1$ for both \Exp and \Erl, $\mu_X = 1.5$ for \ErlExp, $\mu_X = 2/3$ for \ExpErl  and $\/[\xLo, \xUp\/] = \/[0.1, 1.9\/]$ for \UniExp. The comparison between the analysis and the simulation is presented in \Fig{F:simAnalQ} in two scenarios: the left figure shows the results of the setting $\muON = 0.5$ and $\muOFF = 1.5$, which represents a heavy PU traffic model in urban areas, whereas the right figure shows the results of the setting  $\muON = 1.5$ and $\muOFF = 0.5$, which represents  a light PU traffic model in rural areas. Despite the variation in numerical settings,  \Fig{F:simAnalQ} shows that the  analysis results are very similar to the simulation results.

\subsubsection{Equilibrium Convergence} \label{SS:EquiConvg}
We focus on the algorithm and condition for the convergence of the equilibrium joining probability $p_1^*$. We assume that the system operates over successive time periods labeled $t = 0, 1, 2\ldots$. The arrival rate is $\Grl p_1^{t}$ during a period $t$, which is assumed to last long enough for the system to attain the steady state. Since the mean queueing delay varies with the joining probability of the SUs, each type-\Grq SU will make a joining decision  in the next time instant $t+1$ by forming a prediction of the queueing delay denoted by $\hat{T}^{t+1}$. Hence, this SU will join the network at period $t+1$ if and only if $V > \Grq \hat{T}^{t+1} + c_1$. One of possible prediction techniques is every SU expects that the mean queueing delay in the next period is equal to that in the current period: $\hat{T}^{t+1} = \Ex{T( p_1^{t} )}$. Defining $q(p_1^{t})	\triangleq \Fthe \bigl( \Grq_1 ( p_1^{t} ) \bigr)$, we describe the dynamics of SUs' joining probability via two iterative algorithms, namely \emph{static expectations} and  \emph{adaptive expectations} \cite{Evans2001}, of which the SUs' joining probability evolves as follows, respectively
\begin{align}
		p_1^{t+1} = q(p_1^{t})	
\end{align}
and
\begin{align}
		p_1^{t+1} = (1 - \Gra)p_1^{t} + \Gra q(p_1^{t}),	\label{E:adapt}
\end{align}
where $\Gra \in \/(0,1\/]$. We can see that the adaptive method is reduced to the static method when $\Gra = 1$. The static method is also called the naive expectations method \cite{Evans2001} because it assumes that each SU ignores  similar actions of the others. 
In order to alleviate shortcomings of the static expectations, the adaptive expectations method -- with the intuition that only \Gra fraction of SUs make decisions to change at a given time -- allows SUs to learn from and correct for past errors. We obtain the following result which is proved in Appendix~\ref{A:T_Suff}.
\begin{theorem} \label{T:suff}
With any starting point $p_1^{0} \in \/[0, 1\/]$ and an $\Gra \in \/(0,1\/]$, the sufficient condition for the equilibrium convergence of the SUs' joining probability dynamics in \eqref{E:adapt} is 
\begin{align}
	\frac{\ExP{T(1)}}{\Ex{T(1)}} < \frac{1}{\Gra}. \quad \label{E:suff} 
\end{align}
\end{theorem}
If $\Grl < 1/\Ex{\X}$, we can always find a sufficiently small \Gra such that the convergence of the equilibrium is guaranteed \emph{globally}. If $\Grl \geq 1/\Ex{\X}$, this condition is always violated because the left side of \eqref{E:suff} goes to $\infty$. However, this is a sufficient condition, which does not imply that the equilibrium diverges when it is violated. It may still converge \emph{locally} when the starting point is in a neighborhood of the equilibrium.  We will illustrate this effect in the numerical section.

\subsection{Optimal Pricing Mechanisms of \Oo} \label{S:optPrice}
The main focus of this subsection is the operator's use of pricing as a way to maximize its revenue as well as the social welfare of the CR system.
\subsubsection{Revenue-Optimal Pricing} \label{SS:revStrat}
When charging a price $c_1$, \Oo can attain an equilibrium revenue $R_1(c_1) \triangleq c_1\, p_1^*(c_1)$, where $p_1^*(c_1)$ is the equilibrium  at price $c_1$ defined in \eqref{E:fixpoint}. The problem of finding the revenue-optimal price $\cRo$ that maximizes \Oo's equilibrium revenue can then be expressed as
\begin{equation}
\begin{aligned}
& \underset{c_1 \in \left[\cLo, \cUp \right]}{\text{max.}}
& &  R_1(c_1).	\label{E:revOpt} 
\end{aligned}
\end{equation}
Based on \eqref{E:fixpoint}, \eqref{E:waiting}, and \eqref{E:qDelay}, we obtain
\begin{equation} \label{E:pCrev}
    p_1^*(c_1) = \begin{cases}
               0,               &\text{if } c_1 \geq \cUp,	\\
               1,               &\text{if } c_1 \leq \cLo,	\\
               \frac{- \sqrt{\GrP}+ \Ex{\X} \left( \lambda(V-c_1)+\theMax \right)}{\lambda  \theMax \left(2 \Ex{\X}^2-\Ex{\Xsq}\right)},	& \text{otherwise,}
           \end{cases}
\end{equation}
where
\begin{align} 
\cUp &= V,	 \\
\cLo &= \max{\{0, V - \theMax \Ex{T(1)} \}},	 \\
\GrP 	 &= 2 \lambda  \theMax (V-c_1)\Ex{\Xsq}+ \Ex{\X}^2  \Bigl( \lambda (c_1-V)+\theMax \Bigr)^2. 
\end{align}
From \eqref{E:pCrev}, we can see that $R_1(c_1)$ is a concave function; hence, the solution $\cRo$ of the problem \eqref{E:revOpt} can be solved efficiently. When \Oo uses this $\cRo$ for admission pricing and all SUs employ the individual optimal strategy, the corresponding  equilibrium joining probability will be $p_1^*(\cRo)$. 
\subsubsection{Socially-Optimal Pricing} \label{SS:socStrat}
Network social welfare is considered to be the aggregate utility obtained by all SUs. When \Oo charges a price $c_1$, at the equilibrium, only the SUs with $\Grq < \Grq_1(p_1^*(c_1))$ that join the CR network have positive utilities according to \eqref{E:gam1}. Therefore, the network social welfare at price $c_1$ is expressed as follows
\begin{align}
	S_1(c_1) = \int_{0}^{\Grq_1(p_1^*(c_1))} \bigl( V - \Grq \Ex{T(p_1^*(c_1))} \bigr) \, d\Fthe (\Grq),
\end{align}
where $\Grq_1(p_1^*(c_1))$ is the cut-off SU at price $c_1$. The socially-optimal pricing problem can then be cast as
\begin{equation} 
\begin{aligned} \label{E:socOpt}
& \underset{c_1 \in \left[\cLo, \cUp \right]}{\text{max.}}
& &  S_1(c_1).	
\end{aligned}
\end{equation}
However, solving this problem is difficult due to the complex functions $\Grq_1(p_1^*(c_1))$ and $p_1^*(c_1)$. Observing that 
\begin{align}
	p_1^*(c_1)  = \Fthe \bigl( \Grq_1(p_1^*(c_1)) \bigr) = \frac{\Grq_1(p_1^*(c_1))}{\theMax},	\label{E:pCsoc}
\end{align}
we instead change the choice of variable from $c_1$ to a cut-off SU variable denoted by $\theTilO$. Then the new objective function is 
\begin{align}
	S_1(\theTilO) &=  \int_{0}^{\theTilO } \left( V - \Grq \Ex{ T \left(  \theTilO / \theMax \right) } \right) \, d\Fthe (\Grq) =  V \frac{\theTilO}{\theMax}  \, -	\frac{{\theTilO}^2}{2 \theMax} \Ex{ T \left(  \theTilO / \theMax \right) }.	\label{E:S(the)O}
\end{align}
Hence, an equivalent maximization problem of \eqref{E:socOpt} is as follows 
\begin{equation}
\begin{aligned}
& \underset{ \theTilO \geq 0 }{\text{max.}}
& &  S_1 \bigl( \theTilO \bigr).	\label{E:socOpt2}
\end{aligned}
\end{equation}
We can observe that $S_1(\theTilO)$ is concave in its domain; hence, the solution of \eqref{E:socOpt2}, denoted by $\theSo$, can be solved efficiently. Then, from \eqref{E:gam1}, the socially-optimal price $\cSo$ can be calculated as
\begin{align}
	\cSo &= V - \theSo \Ex{T(\theSo / \theMax)}.	\label{E:cSo}	
\end{align}
When \Oo uses this $\cSo$ for admission pricing and all SUs employ the individual optimal strategy, the corresponding  equilibrium joining probability will be $p_1^*(\cSo)$.
\section{Type II: Exclusive-Use Monopoly Market} \label{S:SP2}
In this section, we first investigate the SUs' strategies and the Nash equilibrium. We then examine the \Oe's optimal pricing policies in terms of the revenue and social maximization.
\subsection{SUs' Strategies}
The self-optimizing behaviors of SUs are similar to those in Section~\ref{S:SP1}, where the strategy of each type-$\Grq$ SU is characterized by its joining probability $p(\Grq)$. Given the joining rule $\{p(\Grq), \, \Grq \geq 0\}$, the unconditional probability that a potential arriving SU joins the monopoly \Oe is $\, p_2 = \int_{0}^{\infty} p(\Grq) d \Fthe (\Grq)$. Since a self-optimizing type-\Grq SU will choose $p(\Grq)$ to maximize its expected utility $ p(\Grq) U_2(\Grq) + (1 - p(\Grq))0$, we have the individual optimal strategy of a SU in an exclusive-use monopoly as follows.
\begin{definition}\label{D:IOS2}
An individually-optimizing type-\Grq SU that has $U_2(\Grq)  = V - \Grq \Ex{X} - c_2$ will join \Oe
\begin{itemize}
	\item  with probability $p(\Grq)=1$ if $ U_2(\Grq) > 0 $, which requires
		\begin{align}
			 \Grq < \Grq_2^*, \text{ where } \Grq_2^* \triangleq \frac{V - c_2}{\Ex{X}},  \label{E:gam2}
		\end{align}
	\item with probability $p(\Grq)=0$, otherwise.
\end{itemize}
\end{definition}
Therefore, the equilibrium  of the SUs' joining probability to \Oe is defined as follows.
\begin{definition} \label{D:EJP2} 
$\, p_2^*$ is a Nash equilibrium of SUs' joining probability in an exclusive-use monopoly if it satisfies
\begin{align}
	p_2^* &= \int_{0}^{\infty} p(\Grq) d \Fthe (\Grq)	=  \int_{0}^{ \Grq_2^* } d \Fthe (\Grq)	= \Fthe \bigl( \Grq_2^* \bigr).	\label{E:fixpoint2}
\end{align}
It is clear that, for a given admission price $c_2$, there exists a unique Nash equilibrium of the SUs' joining probability $p_2^*$.
\subsection{Optimal Pricing Mechanisms of \Oe} \label{S:optPrice2}
\subsubsection{Revenue-Optimal Pricing} \label{SS:revStrat2}
When charging a price $c_2$, \Oe can attain an equilibrium revenue $R_2(c_2) \triangleq c_2\, p_2^*(c_2)$, where $p_2^*(c_2)$ is the equilibrium  at price $c_2$ found in \eqref{E:fixpoint2}. It is clear that the revenue-optimal price of \Oe is $\cRe = \frac{V}{2}$, which is the solution of the problem	$\underset{  c_2 \in \left[0, V \right] }{\text{max.}} R_2(c_2)$. 
\end{definition}
\subsubsection{Socially-Optimal Pricing} \label{SS:socStrat2}
The network social welfare  is expressed as follows
\begin{align}
	S_2(c_2) &= \int_{0}^{\Grq_2^* (c_2)} \bigl( V - \Grq \Ex{X} \bigr) \, d\Fthe (\Grq) =  V \frac{\Grq_2^* (c_2)}{\theMax}  \, -	\frac{ \bigl( \Grq_2^* (c_2) \bigr)^2}{2 \theMax} \Ex{X},	\label{E:S(the)E}
\end{align}
where $\Grq_2^* (c_2)$ is a cut-off SU at the price $c_2$ according to \eqref{E:gam2}. It is clear that the  socially-optimal price of \Oe is $\cSe = 0$, which is the solution of the problem $\underset{  c_2 \in \left[0, V \right] }{\text{max.}} S_2(c_2)$. 
%

\section{Type III: Duopoly Market}	\label{S:Duo}
In this section, we consider a duopoly market in a CR network where both \Oo and \Oe compete with each other in terms of pricing in order to maximize their revenues. Based on the prices set by the two operators, a \Grq-type SU will  decide either to join one or to balk to maximize its utility. This duopoly model is illustrated in \Fig{F:duopoly}. The relationship between operators and SUs can be seen as a leader-follower game that can be studied using the two-stage Stackelberg game. Specifically, the operators are the leaders that simultaneously set the prices in Stage I, then SUs will make the joining decisions  in Stage II. 
\subsection{Backward Induction for the Two-Stage Game}
We examine the subgame perfect equilibrium of this Stackelberg game by using a common approach: the backward induction method \cite{Osborne1994,Han2011}. The equilibrium behaviors of the SUs in Stage II will be analyzed first. Then, we investigate how operators determine their prices in Stage I based on the SUs' equilibrium behaviors. 
\subsubsection{SUs' Strategies in Stage II}
In this stage, when two operators are present in the network and set the prices $(c_1, c_2)$, each type-$\Grq$ SU upon arrival will have to choose one of three possible options: join \Oo, join \Oe, or join neither. We denote $p_1(c_1, c_2)$ and $p_2(c_1, c_2)$ as the fraction of SUs  that join \Oo and \Oe, respectively. Henceforth, we simply use the notation $p_1$ and $p_2$. We also define an \emph{indifference SU} as follows
\begin{align}
	\bar{\Grq}(p_1) \triangleq \GamN.
\end{align}

Each SU is assumed to be a rational decision maker in that it only chooses one operator to join if its utility with this operator is both positive and higher than that with the other operator, which corresponds to the following individual optimal strategy.
\begin{definition}\label{D:DuoIOS}
An individually-optimizing type-\Grq SU that has $U_1(\Grq)  = V - \Grq \ETn - c_1$ with \Oo and $U_2(\Grq) = V - \Grq \Ex{X} - c_2$ with \Oe, where $U_1(\bar{\Grq}(p_1)) = U_2(\bar{\Grq}(p_1))$, will join
\begin{itemize}
	\item \Oo with probability $p(\Grq)=1$ if $ U_1(\Grq) > U_2(\Grq) $ and  $ U_1(\Grq) > 0$, which requires
		\begin{align}
			\Grq < \bar{\Grq}(p_1) \text{ and } \Grq < \Grq_1(p_1),	\label{E:DuoGam1}
		\end{align}
	\item \Oe with probability $p(\Grq)=1$ if $ U_2(\Grq) > U_1(\Grq) $ and  $ U_2(\Grq) > 0$, which requires
		\begin{align}
			 \bar{\Grq}(p_1) < \Grq  < \Grq_2^*,	\label{E:DuoGam2}
		\end{align}
	\item neither with probability $p(\Grq)=1$ if $ U_1(\Grq) < 0 $ and  $ U_2(\Grq) < 0$, which requires
		\begin{align}
			\Grq > \Grq_1(p_1) \text{ and } \Grq > \Grq_2^*.	\label{E:DuoGam3}
		\end{align}
\end{itemize}
\end{definition}
Recall that $\Grq_1(p_1)$ and $\Grq_2^*$ are given in \eqref{E:gam1} and \eqref{E:gam2}, respectively. Based on \eqref{E:DuoGam1}, \eqref{E:DuoGam2} and \eqref{E:DuoGam3}, the  unconditional joining probabilities $(p_1, p_2)$ are as follows 
	\begin{align}
		\left(p_1, \, p_2\right) = \left(  \int_{0}^{\min \{\Grq_1(p_1), \, \bar{\Grq}(p_1) \} } d \Fthe (\Grq), \,  \int_{\bar{\Grq}(p_1)}^{\Grq_2^*} d \Fthe (\Grq) \right).	\label{E:DuoP}
	\end{align}

With the time slot model as in Subsection~\ref{SS:EquiConvg} and according to \eqref{E:DuoP}, the SUs' joining probability dynamics in the duopoly can be described as follows
\begin{enumerate}[(a)]
	\item If $\bar{\Grq}\bigl( p_1^{t} \bigr) < 0$, which leads to $c_2 < c_1$ since $\ETt > \Ex{X}, \; \forall  t$, then \Oo is eliminated from the competition, leaving \Oe as a monopoly. We have
		\begin{align}
			\left(p_1^{t+1}, \, p_2^{t+1}\right) = \Bigl( 0, \; \Fthe \left( \Grq_2^* \right) \Bigr), \forall  t. 	\label{E:DuoDyn1}
		\end{align}
	\item If $ 0 < \bar{\Grq}\bigl( p_1^{t} \bigr) < \Grq_2^*$, which leads to $ \Grq_2^* < \Grq_1(p_1^{t}) $, then we have 
		\begin{align}
			\left(p_1^{t+1}, \, p_2^{t+1}\right) = \Bigl( \Fthe \bigl( \bar{\Grq}\bigl( p_1^{t} \bigr) \bigr), \, \Fthe \left( \Grq_2^* \right) - \Fthe \bigl( \bar{\Grq}\bigl( p_1^{t} \bigr) \bigr) \Bigr). 	\label{E:DuoDyn2}
		\end{align}
	\item If $ \bar{\Grq}\bigl( p_1^{t} \bigr) > \Grq_2^*$, which leads to $ \Grq_2^* > \Grq_1(p_1^{t}) $, then \Oe is eliminated from the competition, leaving \Oo as a monopoly. We have
		\begin{align}
			\left(p_1^{t+1}, \, p_2^{t+1}\right) = \Bigl( \Fthe \bigl( \Grq_1 \bigl( p_1^{t} \bigr) \bigr), \, 0 \Bigr).	\label{E:DuoDyn3}
		\end{align}
\end{enumerate}

Since there exists a sufficiently small $\Gre$ and the corresponding $p^{\Gre} \triangleq \min\{1, \frac{1}{\Grl \EX} - \Gre \}$  such that 
\begin{align}
 	p_1^{t} &\leq  p^{\Gre}, 	\forall  t,	\label{E:pEps}
\end{align}
we define an equilibrium  $(p_1^*, p_2^*)$ in a duopoly market with the given prices $(c_1, c_2)$ as follows. 
\begin{definition} \label{D:DuoEJP} 
Given a sufficient small $\Gre$, $(p_1^*, p_2^*)$ is a Nash equilibrium  of SUs' joining probability in a duopoly market if it satisfies
	\begin{align} \label{E:DuoFixpoint}
		&(p_1^*, p_2^*) = \begin{cases}
	   		\Bigl( 0, \; \Fthe \left( \Grq_2^* \right) \Bigr),       &\text{ if } \theBarLo < 0,	\\
	        \Bigl( \Fthe \left( \bar{\Grq}(p_1^*) \right), \; \Fthe \left( \Grq_2^* \right) - \Fthe \left( \bar{\Grq}(p_1^*) \right) \Bigr),       &\text{ if } 0 < \theBarLo < \Grq_2^*,	\\
	   		\Bigl( \Fthe \left( \Grq_1(p_1^*) \right), \; 0 \Bigr),       &\text{ if } \theBarLo > \Grq_2^*.
	   \end{cases}
	\end{align}
\end{definition}
We have the following result which is proved in Appendix~\ref{A:T_DuoUniq}.
\begin{theorem} \label{T:DuoUniq}
For a given admission price pair $(c_1, c_2)$, there exists a unique Nash equilibrium  of the SUs' joining probability $(p_1^*, p_2^*)$ satisfying \eqref{E:DuoFixpoint} in a duopoly market.
\end{theorem}

We can see that both  $ \theBarLo < 0 $ and $ \theBarLo > \Grq_2^* $ correspond to the equilibrium behaviors of the shared-use and exclusive-use monopolies in Section~\ref{S:SP1} and Section~\ref{S:SP2}, respectively. Therefore, we will focus only in the case  $0 < \theBarLo < \Grq_2^*$ in what follows.

Regarding the convergence of the equilibrium $(p_1^*, p_2^*)$, the static expectations method is presented in \eqref{E:DuoDyn2}. The adaptive expectations method can also be presented as follows
\begin{align}
		\mathbf{p}^{t+1} = (1 - \Gra)\mathbf{p}^{t} + \Gra \mathbf{q_d}(\mathbf{p}^{t}),	\label{E:DuoAdapt}
\end{align}
where $\Gra \in \/(0,1\/]$, $\mathbf{p}^{t} = \left(p_1^{t}, \, p_2^{t} \right)$ and
\begin{align}
 \mathbf{q_d}(\mathbf{p}^{t}) 
 							  &\triangleq \Bigl( \Fthe \bigl( \bar{\Grq}\bigl( p_1^{t} \bigr) \bigr), \, \Fthe \left( \Grq_2^* \right) - \Fthe \bigl( \bar{\Grq}\bigl( p_1^{t} \bigr) \bigr) \Bigr).	\label{E:qDuoDef}
\end{align}

We obtain the following result which is proved in Appendix~\ref{A:T_DuoSuff}.
\begin{theorem} \label{T:DuoSuff}
With any starting point $(p_1^{0}, p_2^{0}) \in [0, 1]^2$ and an $\Gra \in \/(0,1\/]$, the sufficient condition for the equilibrium convergence of the SUs' joining probability dynamics \eqref{E:DuoAdapt} is 
\begin{align}
	\frac{\ExP{T( 1 )}}{\Ex{T(1)} -\Ex{X}} < \frac{1}{\Gra}.	\label{E:DuoSuff}
\end{align}
\end{theorem}
Since the left side of \eqref{E:DuoSuff} is strictly greater than that of \eqref{E:suff}, we see that the equilibrium convergence condition of a duopoly in \eqref{E:DuoSuff} is more stringent than that of the shared-use monopoly in \eqref{E:suff}.
\subsubsection{Price Competition in Stage I}
In this stage, the operators determine their pricing strategies based on $(p_1^*, p_2^*)$ in Stage II. Given a pair of prices $(c_1, c_2)$, the equilibrium revenue of the operator $i$ is
\begin{align}
	R_i(c_1, c_2) = c_i p_i^*, \quad i = 1, 2.
\end{align}
Here, $(p_1^*, p_2^*)$ is given in \eqref{E:DuoFixpoint} in the case of the duopoly coexistence with $0 < \theBarLo < \Grq_2^*$, which corresponds to the condition
\begin{align}
	c_1 < c_2 < u(c_1), 	\label{E:PriceCond}
\end{align}
where 
\begin{align}
	u(c_1) \triangleq \frac{ V\bigl (\Ex{T(p^{\Gre})} -\Ex{X} \bigr) + \Ex{X} c_1 }{ \Ex{T(p^{\Gre})} }.
\end{align}

The competition between two operators in Stage I can then be modelled as the following game
\begin{itemize}
	\item Players: \Oo and \Oe,
	\item Strategy: \Oo chooses price $c_1 \in (0, c_2)$; \Oe chooses price $c_2 \in (0, u(c_1))$,
	\item Payoff function: 	$R_i(c_1, c_2), \; i = 1, 2.$
\end{itemize}
We denote the Stage I game equilibrium by $(c_1^*, c_2^*)$, and define $p_i^{\circledast} \triangleq p_i^*(c_1^*,c_2^*),\; i =1, 2$.  
\begin{theorem} \label{T:DuoNash}
There exists a unique Nash equilibrium of a Stage I game such that
\begin{equation} \label{E:DuoNash}
(c_1^*, c_2^*) = \left(\frac{ V \bigl( \Ex{T(p_1^{\circledast})} - \Ex{X} \bigr) }{4\Ex{T(p_1^{\circledast})} - \Ex{X}},  \frac{2 V \bigl( \Ex{T(p_1^{\circledast})} - \Ex{X} \bigr) }{4\Ex{T(p_1^{\circledast})} - \Ex{X}} \right) 
\end{equation}
where 
\begin{align}
	p_1^{\circledast} = \frac{2 V}{\sqrt{\GrD} + \EX \left( \lambda  V +3 \theMax \right)} 
	\text{ with } \GrD = \Ex{\X}^2 (\Grl V - 3\theMax )^2 + 8\theMax \lambda V \Ex{\Xsq}.	\label{E:DuoPII}
\end{align}
\end{theorem}

The proof of Theorem~\ref{T:DuoNash} is given in Appendix~\ref{A:T_DuoNash}. We then examine whether $(c_1^*, c_2^*)$ satisfies the condition \eqref{E:PriceCond} or not. Since the lower bound of \eqref{E:PriceCond} is clearly satisfied, we check the upper bound condition $c_2^* < u(c_1^*)$, which is equivalent to the following inequality after some algebra manipulations
\begin{align}
	2 \Ex{T(p_1^{\circledast})}  \Ex{T(p^{\Gre})} > \Ex{X} \bigl( 3 \Ex{T(p_1^{\circledast})} - \Ex{T(p^{\Gre})} \bigr). 	\label{E:UpCond}
\end{align}
It turns out that \eqref{E:UpCond} is true since $\Ex{T(p^{\Gre})} > \Ex{X}$ and $\Ex{T(p^{\Gre})} \geq  \Ex{T(p_1^{\circledast})}$.
\subsection{Equilibrium Summary}
 We summarize all equilibrium cases in Table~\ref{Tab:Sum}. Since \Oe has dedicated channels for SUs and provides less delay than that of \Oo, it is intuitive that \Oe becomes a monopolist when $c_1 > c_2$. However, if \Oe's price is much higher than \Oo's in case of $c_2 > u(c_1)$, then \Oo becomes a monopolist. Only in the case of $ c_1 < c_2 < u(c_1)$, both operators can share the market and the unique subgame perfect equilibrium of the Stackelberg game is $\bigl( (c_1^*, c_2^*),\; (p_1^{\circledast}, p_2^{\circledast}) \bigr)$. To attain this equilibrium, the operators first update the statistical information of SUs and PUs to calculate $(c_1^*, c_2^*)$ according to \eqref{E:DuoNash} and broadcast. Based on these prices, the SUs then employ the individual optimal strategy, inducing the result $(p_1^{\circledast}, p_2^{\circledast})$.
\begin{table*}[!t]
\renewcommand{\arraystretch}{1.3}
\caption{Equilibrium Summary} \label{Tab:Sum} \centering
\begin{tabular}{c||c|c|c}
\hline Pricing space & $c_1 > c_2$ & $ c_1 < c_2 < u(c_1)$ & $c_2 > u(c_1)$  \\
\hline Equilibrium & \parbox[t]{4.05cm}{\Oe monopoly Nash equilibrium \\  $(p_1^*, p_2^*) = 
	   		\bigl( 0, \; \Fthe \left( \Grq_2^* \right) \bigr)$} & \parbox[t]{4.22cm}{Duopoly subgame perfect \\ equilibrium $\bigl( (c_1^*, c_2^*),\; (p_1^{\circledast}, p_2^{\circledast}) \bigr)$} & \parbox[t]{4.05cm}{\Oo monopoly Nash equilibrium \\ $(p_1^*, p_2^*) = \bigl( \Fthe \left( \Grq_1(p_1^*) \right), \; 0 \bigr)$}  \\
\hline
\end{tabular}
\end{table*}

\section{Numerical Results} \label{S:numResult}
In this section, we apply the analysis results to numerically illustrate the SUs' equilibrium behaviors and the optimal pricing strategies in the \Oo monopoly market first, and then the interaction between \Oo and \Oe in the duopoly market. To facilitate the illustration, the parameter settings adhere to the following order of \Exp, \Erl, \UniExp, \ErlExp and \ExpErl examples with light PU traffic model (i.e. $\muON = 1.5$ and $\muOFF = 0.5$) in Section~\ref{SSS:settings}. Furthermore, we set $V=1$, $\theMax =1$, $\Gra = 0.3$, and $\Gre = 0.01$.
\begin{figure}[!t]
\centering
\includegraphics[width=0.67\textwidth]{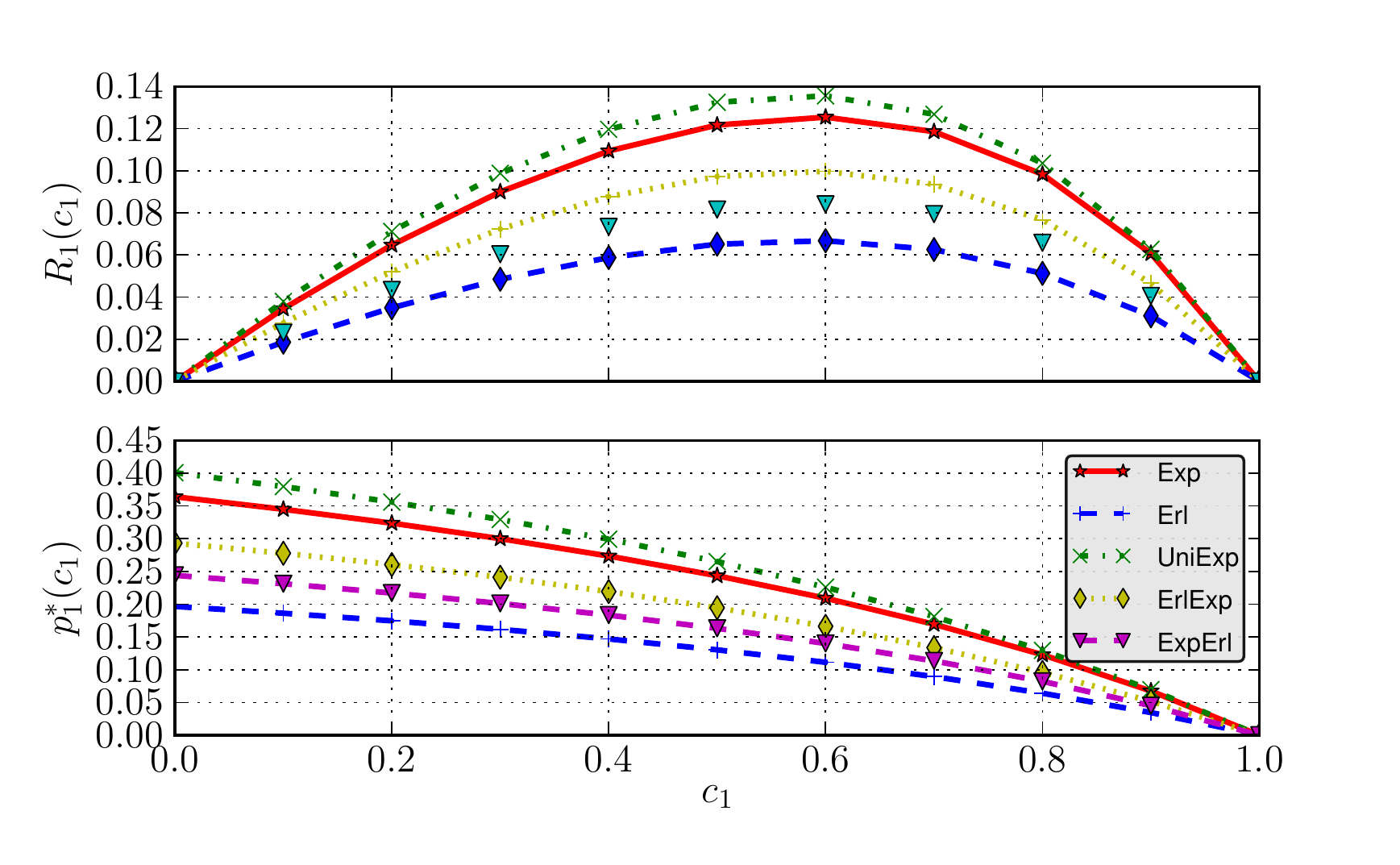}
\caption{ \Oo's revenues (top plot) and SUs' equilibrium joining probabilities (bottom plot) as functions of the admission price.  }
\label{F:rev}
\end{figure}
\begin{figure}[!t]
\centering
\includegraphics[width=0.67\textwidth]{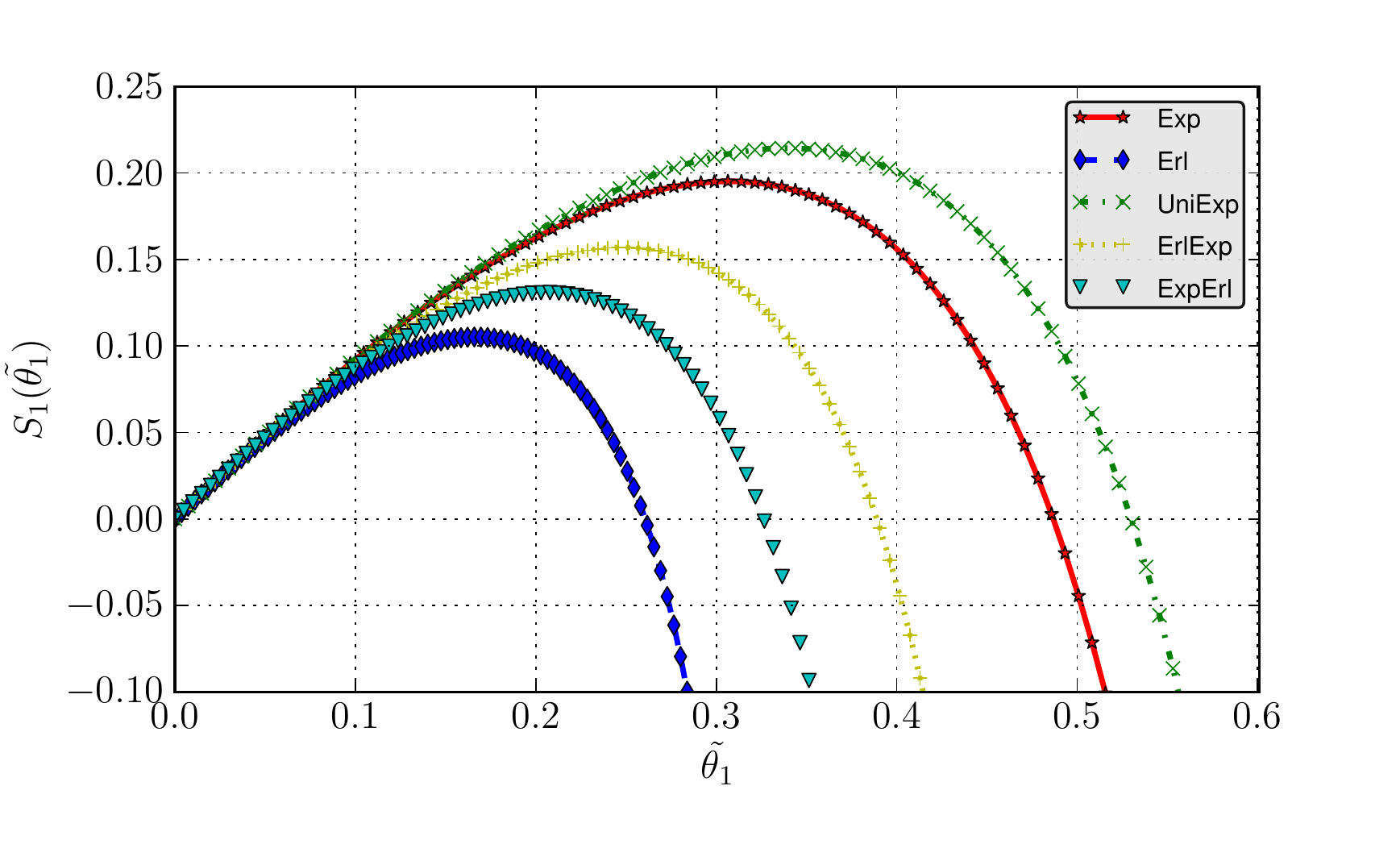}
\caption{ The network social welfare as  functions of the cut-off SUs. }
\label{F:soc}
\end{figure}
\begin{figure}[!t]
\centering
\includegraphics[width=0.67\textwidth]{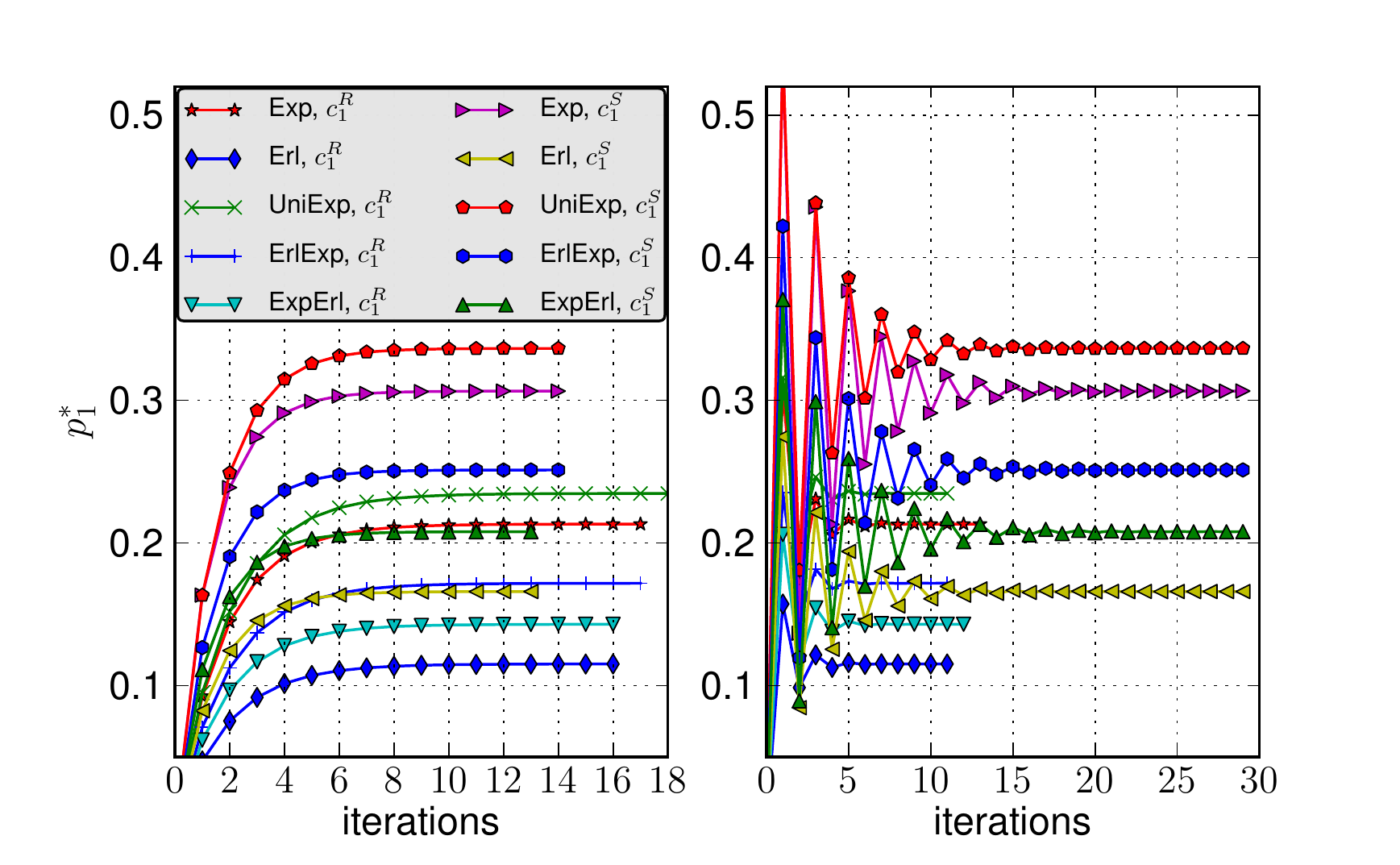}
\caption{ The convergence of equilibrium joining probabilities $p_1^*(\cRo)$ and $p_1^*(\cSo)$ with static expectations (left plot) and adaptive expectations (right plot). }
\label{F:SP1}
\end{figure}
\begin{figure}[!t]
\centering
\includegraphics[width=0.67\textwidth]{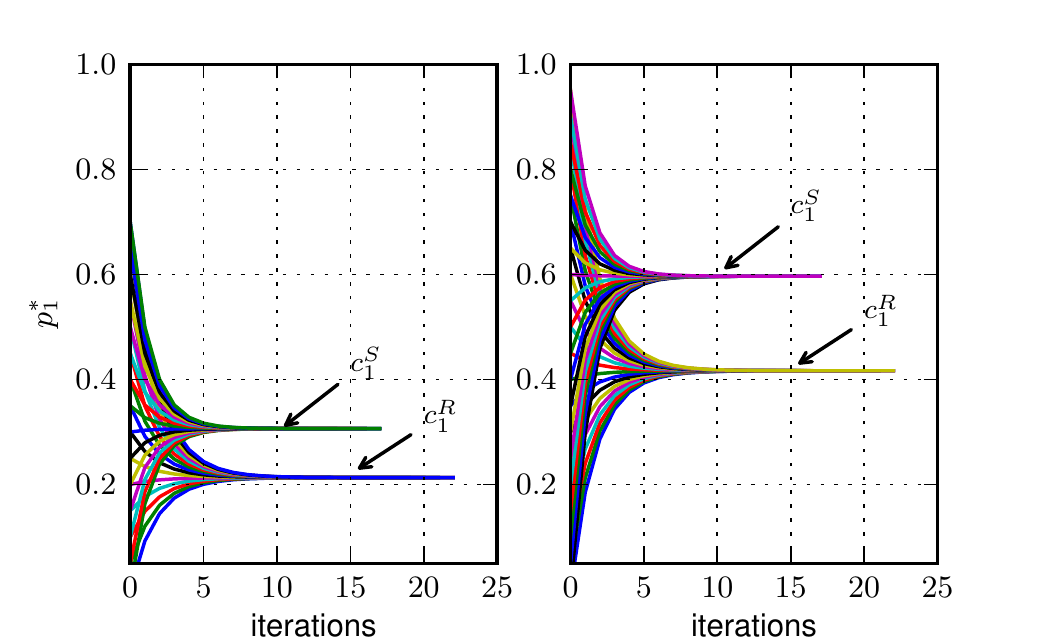}
\caption{ $p_1^*(\cRo)$ and $p_1^*(\cSo)$ converge locally (left plot) and globally (right plot) when the condition in \eqref{E:suff} is violated and satisfied, respectively. }
	\label{F:convg}
\end{figure}
\begin{figure}[!t]
\centering
	\subfloat[$ c^*_1, \, c^*_2$]{\includegraphics[width=0.63\textwidth]{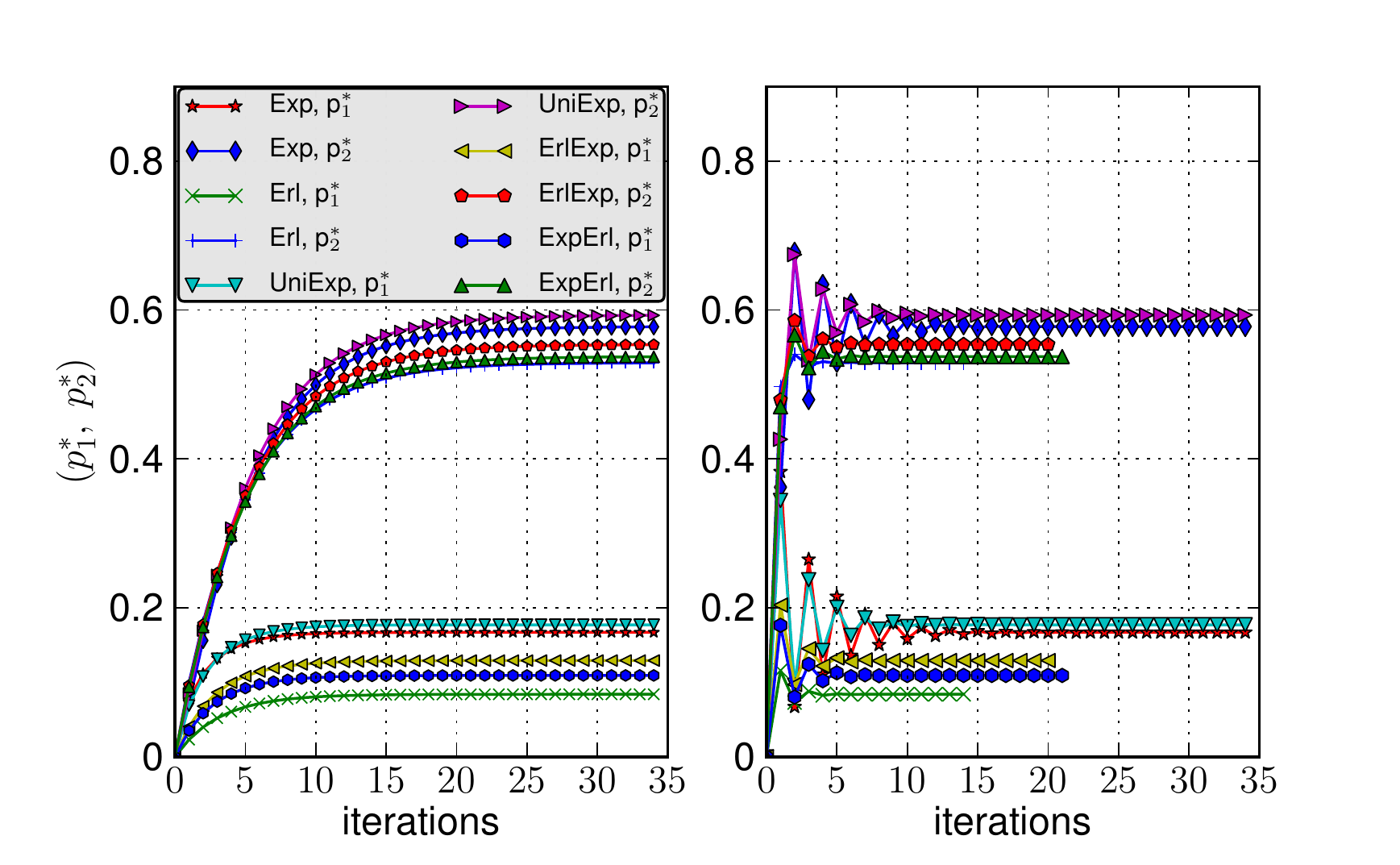}
	\label{F:duo}}
	\hfil
	\subfloat[$ c_1 = \cRo, \, c_2 = 0.99 \, u(c_1) $]{\includegraphics[width=0.63\textwidth]{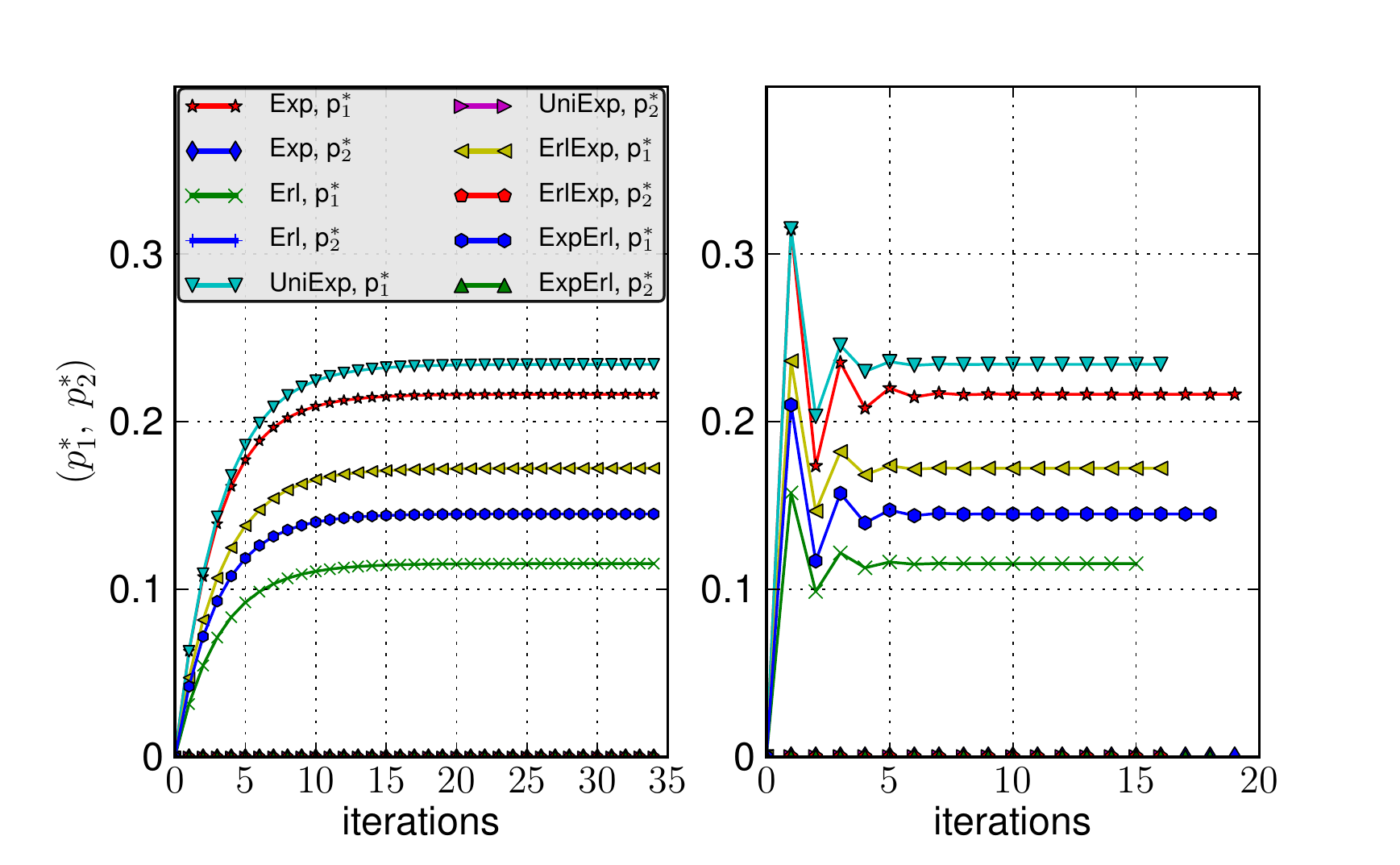}
	\label{F:monoSP1}}
	\hfil
	\subfloat[$ c_1 = 0.99\,c_2, \, c_2 = 0.5 $]{\includegraphics[width=0.63\textwidth]{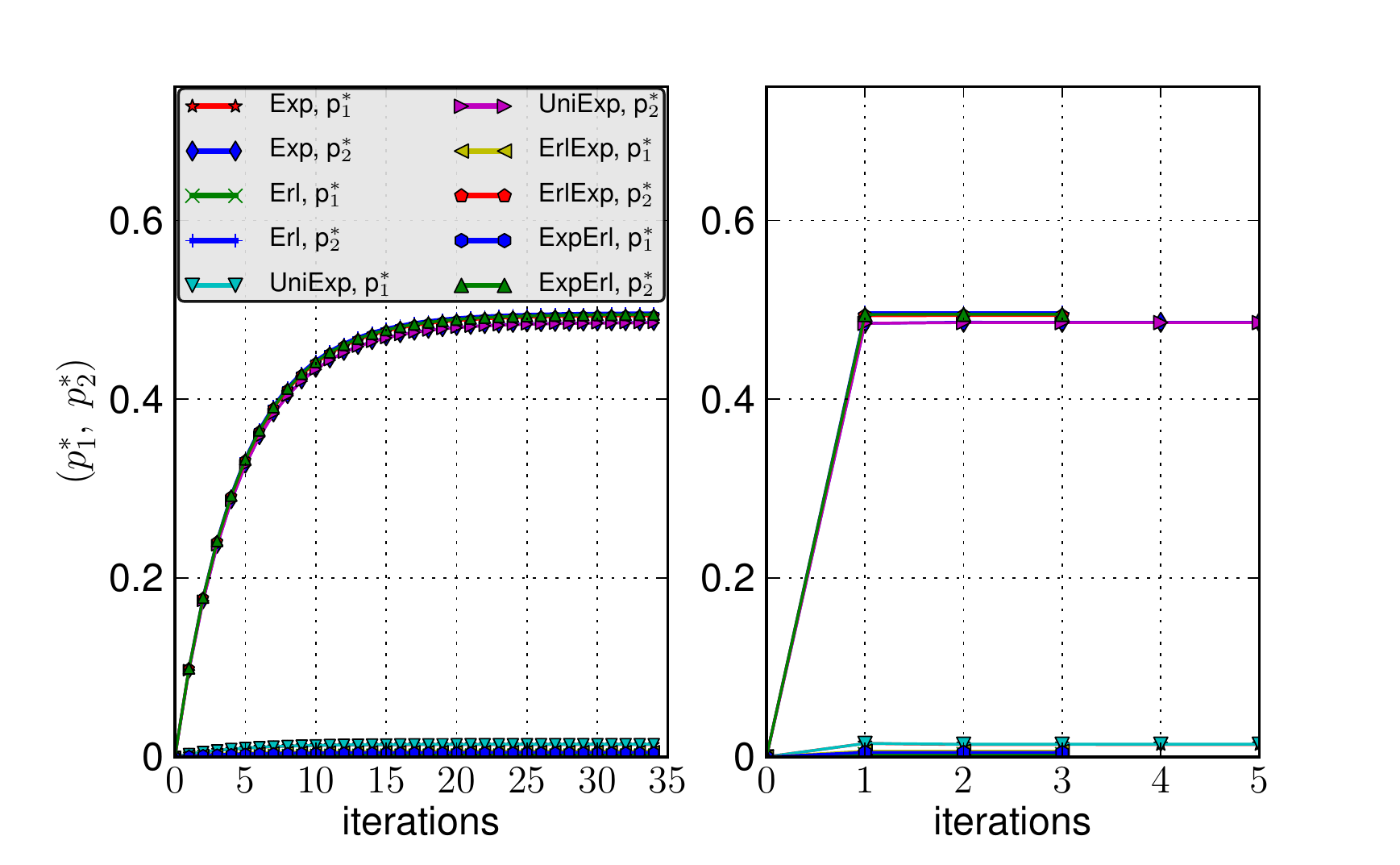}
	\label{F:monoSP2}}
	\caption{Convergence behaviors of the duopoly equilibrium $(p_1^*, p_2^*)$ with three different price pair settings.}
	\label{F:pricePair}
\end{figure}
\subsection{Shared-Use Monopoly}

\subsubsection{Revenue Optimization}
The top part of \Fig{F:rev} shows graphs of \Oo's revenue with respect to price $c_1$. We can see that the revenue functions have convex forms and their maximum values are achieved nearly at the same price ($0.58$) with the corresponding revenues $0.13, 0.07, 0.14, 0.1$ and $0.08$ with respect to the order of the example settings. At $c_1=0$ and $c_1 = V =1$, all revenues are zero, which is clear due to the revenue function and the individual optimal strategy definitions. The equilibrium joining probability $p_1^*(c_1)$ is plotted in the bottom part of \Fig{F:rev}. At the price $\cRo = 0.58$, we can see that the corresponding $p_1^*(\cRo)$ are $0.21, 0.11, 0.23, 0.17$ and $0.14$ with respect to the order of example settings. This plot also demonstrates that when $c_1$ is increased, $p_1^*(c_1)$ is decreased.   

\subsubsection{Social Optimization}
The network social welfare as a function of cut-off user $\theTilO$ is shown in \Fig{F:soc}. It can be seen that the socially-optimal cut-off SUs $\theSo$ are  $0.3, 0.16, 0.33, 0.25$ and $0.21$, and the corresponding socially-optimal values $S_1(\theSo )$ are $0.19, 0.1, 0.21, 0.16$ and $0.13$ with respect to the order of the example settings. The respective socially-optimal prices $\cSo$ can be calculated according to \eqref{E:cSo}. Compared with the bottom plot of \Fig{F:rev}, these prices map correctly with the corresponding values of $p_1^*(\cSo)$, which is equal to $\theSo$ since $\theMax = 1$.

\subsubsection{Equilibrium Convergence Dynamics}
With the starting point $p_1^{0}$ set to zero, the equilibrium convergence of all settings using static and adaptive expectations are illustrated in \Fig{F:SP1}. Although the condition in \eqref{E:suff} is violated in all five settings (i.e. $\Grl > 1/\Ex{\X}$), it can still be seen that all joining probabilities converge to the expected equilibrium points presented previously as Theorem~\ref{T:suff} gives a sufficient but not necessary condition. 

In \Fig{F:convg}, we examine the local and global convergence when the condition in \eqref{E:suff} is violated and satisfied, respectively.  With the \Exp setting, the left plot of \Fig{F:convg} shows that the equilibria $p_1^*(\cRo)$ and $p_1^*(\cSo)$ can converge if we choose the starting points in the range of $[0, 0.75]$. We observe that if the starting point is larger than $0.76$, the divergence occurs. The right plot of \Fig{F:convg} shows the global convergence of the equilibria $p_1^*(\cRo)$ and $p_1^*(\cSo)$ in a new setting that is the same as the \Exp setting except that the parameter $\mu_X$ is changed to 2, inducing the condition in \eqref{E:suff} is satisfied with \Gra = 0.3. 

\subsection{Duopoly}
We continue to illustrate the price competition between \Oo and \Oe. We first consider the effect of the Stage I's equilibrium $(c_1^*, c_2^*)$ on the Stage II's equilibrium $(p_1^*, p_2^*)$. From \eqref{E:DuoNash}, the equilibrium values are $(c_1^*, c_2^*) = (0.13, 0.26), (0.19, 0.38), (0.11, 0.22), (0.16, 0.32)$ and $(0.18, 0.36)$ with respect to the order of example settings. With these equilibrium prices, the corresponding $(p_1^*, p_2^*)$ convergence of all settings is shown in \Fig{F:duo}.

We next illustrate the tendency of a duopoly to form a monopoly if the condition in \eqref{E:PriceCond} is violated. We first choose a price pair that is close to the upper bound of this condition, where $c_1 = \cRo = 0.58$ and $ c_2 = 0.99\, u(c_1)$. \Fig{F:monoSP1} shows that with this price pair, the $p_2^*$ of all settings converges closely to 0, whereas $p_1^*$ converges closely to the equilibrium in the \Oo monopoly shown in \Fig{F:rev}. We then choose a price pair that is close to the lower bound of the condition in \eqref{E:PriceCond}, where $c_2$ is set arbitrarily to 0.5  and $ c_1 = 0.99\, c_2$. With this price pair, as shown in \Fig{F:monoSP2}, the $p_1^*$ of all settings converges closely to 0, whereas $p_2^*$  of all settings converges closely to $\Fthe \left( \Grq_2^* \right) = \frac{0.5}{\Ex{X}}$, the equilibrium joining probability of the \Oe monopoly.

The data shown in \Fig{F:pricePair} not only validate our analysis of the price competition of a duopoly, but also provide the convergence behaviors of different methods. In \Fig{F:pricePair}, all the graphs in the left column show the convergence with the adaptive expectations method, whereas the graphs in the right column show the convergence with the static expectations method.

\section{Conclusion} \label{S:conclusion}
This paper describes the price-based spectrum access control between the operators and SUs in three market scenarios. The interactions between the first monopolist operator with shared-use DSA and delay-sensitive SUs is examined through a queueing analysis representing the SU's congestion due to the shared single channel. We show that there exists a unique Nash equilibrium in a non-cooperative game where the SUs are the players employing individual optimal strategies for spectrum access. We also provide a sufficient condition and the iterative algorithms for the equilibrium convergence. The pricing mechanisms of the operators are also considered for two problems of revenue and social welfare maximization. The second monopolist operator using exclusive-use DSA has many channels to dedicate to SUs. Owing to the separate channels, the analysis of the interactions between the operator and the SUs is straightforward, yet provides  useful insights in the third market analysis. In the third duopoly market, we study the price competition between two operators using shared-use and exclusive use DSA. We formulate this competition as a two-stage Stackelberg game. The equilibrium behaviors of the SUs in Stage II are analyzed first, and we then examine how operators determine their prices in Stage I based on the SUs' behaviors. Using the backward induction method, we show that there exists a unique equilibrium in this game and investigate the equilibrium convergence.
 

\appendices
\section{Proof of Theorem \ref{T:Uniq}} \label{A:T_Uniq}
We first show the existence and uniqueness of the equilibrium. Defining $\Phi(p_1) \triangleq \Fthe \left( \Grq_1(p_1) \right) - p_1$ with $p_1 \in \/[0,1\/]$, we can see that $\Phi(p_1)$ is a strictly decreasing function because $\Fthe(.)$ is an increasing function and  $\Grq_1(p_1)$ is a  strictly decreasing function (since $\Ex{T(p_1)}$ is strictly increasing) on their domains. By Definition~\ref{D:EJP}, $p_1^*$ is an equilibrium if and only if it is a root of $\Phi(p_1)$. Hence, it suffices to show that $\Phi(p_1)$ has a unique root on its domain as follows.\\
When $V \leq c_1 $, we clearly see that $p_1^* = 0$ is the unique root of $\Phi(p_1)$.\\
When $V \geq c_1 + \theMax \Ex{T(1)} $, we clearly see that $p_1^* = 1$ is the unique root of $\Phi(p_1)$.\\
When $c_1  < V < c_1 + \Ex{T(1)} \theMax $,  we have two following  cases
\begin{enumerate}[(a)]
\item There exists a $p_1' \in (0, 1)$ such that 
	\begin{align}
		c_1 < V = c_1 + \Ex{T( p_1')} \theMax < c_1 + \Ex{T( 1 )} \theMax \label{F:p'},
	\end{align}
since $\Ex{T(p_1)}$ is a strictly increasing function. Then we observe that
	\begin{align}
	\Phi(p_1) &= 1 - p_1 > 0, \quad \forall p_1 \in \/[0, p_1'\/],	\label{F:phi1}\\
	\Phi(1) &= \Fthe \left( \Grq_1(1) \right) - 1 < 0. \label{F:phi2}
	\end{align}
With \eqref{F:phi1}, \eqref{F:phi2} and the fact that  $\Phi(p_1)$ is a continuous and strictly decreasing function, we see that $\Phi(p_1)$ has a unique root $p_1^* \in (p_1', 1)$ in this case.
\item If there does not exist any $p'$ satisfying \eqref{F:p'}, then we have $c_1  < V < c_1 + \Ex{T(p_1)} \theMax, \; \forall p_1 \in \/[0, 1\/].$ We observe that 
	\begin{align}
		\Phi(0) &= \Fthe \left( \Grq_1(0) \right) > 0, \label{F:phi1'}\\
		\Phi(1) &= \Fthe \left( \Grq_1(1) \right) - 1 < 0. \label{F:phi2'}
	\end{align}
With \eqref{F:phi1'}, \eqref{F:phi2'} and the fact that  $\Phi(p_1)$ is a continuous and strictly decreasing function, we see that $\Phi(p_1)$ has a unique root $p_1^* \in (0, 1)$ in this case.
\end{enumerate}

Next, we show that this unique root is a Nash equilibrium. When all SUs experience the same mean queueing delay $\Ex{T(p_1^{*})}$, $p_1^*$ will be a Nash equilibrium if no SU of any type \Grq can increase its utility by choosing an entrance probability different from $p(\Grq)$ in Definition~\ref{D:IOS1}. To see this, consider a specific type-\Grq SU 
	
\begin{enumerate}[(a)]
		\item If $V > \Grq \Ex{T(p_1^{*})} + c_1$, according to Definition~\ref{D:IOS1}, this SU will join with probability $p(\Grq)=1$; hence, its expected utility is $V - \Grq \Ex{T(p_1^{*})} - c_1 > 0$. If this SU deviates from this individual optimal strategy by choosing another strategy $0 \leq \tilde{p}(\Grq) < 1$, it will receive an expected utility $\tilde{p}(\Grq)(V - \Grq \Ex{T(p_1^{*})} + c_1) < V - \Grq \Ex{T(p_1^{*})} - c_1.$	Therefore, such a SU has no incentive to deviate from its current strategy. Conversely, if such a SU chooses another strategy $\tilde{p}(\Grq)<1$, it will find that it can increase its expected utility by switching to $p(\Grq)=1$.
		\item If $V < \Grq \Ex{T(p_1^{*})} + c_1$,  by deviating from the individual optimal strategy (i.e. $p(\Grq)=0$), the SU will receive a strictly smaller expected utility; hence, this SU has no incentive to deviate from its current strategy.
		\item If $V = \Grq \Ex{T(p_1^{*})} + c_1$, by deviating from the individual optimal strategy, the expected utility of this SU will still be zero; hence, this SU also has no incentive to deviate from its current strategy.
\end{enumerate}

\section{Proof of Theorem \ref{T:suff}} \label{A:T_Suff}
Since $q(p_1):[0, 1]\mapsto[0, 1]$ is differentiable, according to the contraction mapping \cite{Bertsekas1989}, the equilibrium  can be achieved and is stable for any starting point $p_1^{0} \in [0, 1]$ if the following condition is satisfied
\begin{align}
\Gra \bigl\vert q'(p_1) \bigr\vert < 1,	\quad	\forall p_1 \in \/[0, 1\/].	\label{E:suff1}
\end{align}
With $q(p_1)	= \Fthe \bigl( \Grq_1 ( p_1 ) \bigr)$, we have
\begin{align}
\bigl\vert q'(p_1) \bigr\vert = \Biggl\vert - \fthe \bigl( \Grq_1 ( p_1 ) \bigr) \Grq_1 ( p_1 ) \frac{\ExP{T(p_1)}}{\Ex{T(p_1)}}\Biggr\vert	&\leq \Biggl\vert \max_{\Grq \in \/[0,\theMax\/]}\fthe(\Grq) \Grq \Biggr\vert 
		  	  \Biggl\vert \max_{p_1 \in \/[0,1\/]} \frac{\ExP{T(p_1)}}{\Ex{T(p_1)}} \Biggr\vert	= \frac{\ExP{T(1)}}{\Ex{T(1)}}, \label{E:suff2}
\end{align}
where the third equality can be determined because $\max_{\Grq \in \/[0,\theMax\/]}\fthe(\Grq)\, \Grq = 1$ and $\frac{\ExP{T(p_1)}}{\Ex{T(p_1)}}$ is a positive and increasing function, which attains the maximum value at the upper boundary point $p_1 = 1$. From \eqref{E:suff1} and \eqref{E:suff2}, we complete the proof.

\section{Proof of Theorem \ref{T:DuoUniq}} \label{A:T_DuoUniq}
If $\theBarLo < 0$, from \eqref{E:DuoDyn1}, the unique Nash equilibrium is $\bigl( 0, \, \Fthe \left( \Grq_2^* \right) \bigr)$ corresponding to the case of exclusive-use monopoly analyzed in Section \ref{S:SP2}.

If $ \theBarLo > \Grq_2^* $, since we have $ \bar{\Grq}(p^{t}) > \theBarLo > \Grq_2^*, \;\forall  t$ according to \eqref{E:pEps}, the unique Nash equilibrium is $ \bigl( \Fthe \left( \Grq_1(p_1^*) \right), \, 0 \bigr)$ corresponding to the case of shared-use monopoly analyzed in Section \ref{S:SP1}.

If $ 0 < \theBarLo < \Grq_2^* $, we first focus on the existence and uniqueness of an equilibrium as in Definition~\ref{D:DuoEJP}. We define $\Omega(p_1) \triangleq \Fthe \left( \bar{\Grq}(p_1) \right) - p_1$ for $p_1 \in \/[0,1\/]$. We can see that $\Omega(p_1)$ is a strictly decreasing function  because $\Fthe(.)$  and  $\bar{\Grq}(p_1)$ are increasing  and  strictly decreasing functions, respectively, on their domains. By Definition~\ref{D:DuoEJP}, $p_1^*$ is an equilibrium  if and only if it is a root of $\Omega(p_1)$. Hence, it suffices to show that $\Omega(p_1)$ has a unique root on its domain. Based on the fact that 
\begin{align}
	\Omega(0) &= \Fthe \left(\bar{\Grq}(0) \right) > 0, \\
	\Omega(1) &= \Fthe \left( \bar{\Grq}(1) \right) - 1 < 0,
\end{align}
and $\Omega(p_1)$ is a continuous and strictly decreasing function, we see that $\Omega(p_1)$ has a unique root $p_1^* \in (0, 1)$.
Since $p_2^* = \Fthe \left( \Grq_2^* \right) - \Fthe \left( \bar{\Grq}(p_1^*) \right)$ only depends on the unique $p_1^*$, it is clear that $(p_1^*, p_2^*)$ is a unique equilibrium  of Definition~\ref{D:DuoEJP}. We next present that this $(p_1^*, p_2^*)$ is a Nash equilibrium by showing that at this point,  no SU of any type \Grq can increase its utility by deviating from the individual optimal strategy in Definition~\ref{D:DuoIOS}. We note that $\bar{\Grq}(p_1^*) < \Grq_2^*$ because if $\bar{\Grq}(p_1^*) > \Grq_2^*$, \eqref{E:DuoDyn3} shows a contradiction. Therefore, with $0 < \bar{\Grq}(p_1^*) < \Grq_2^*$, we consider a specific type-\Grq SU 

\begin{enumerate}[(a)]
	\item If $0 < \Grq < \bar{\Grq}(p_1^*) < \Grq_2^* $, we have $V - \Grq \ET - c_1 > V - \Grq \Ex{X} - c_2$. Moreover, $\bar{\Grq}(p_1^*) < \Grq_2^*$ leads to $\bar{\Grq}(p_1^*) < \Grq_1(p_1^*)$, which shows that $V - \Grq \ET - c_1 > 0$. Therefore, this SU has no intention to deviate from the individual optimal strategy defined in Definition~\ref{D:DuoIOS}.
	\item If $0 <  \bar{\Grq}(p_1^*) < \Grq < \Grq_2^*$, we have $V - \Grq \ET - c_1 < V - \Grq \Ex{X} - c_2$ and $V - \Grq \Ex{X} - c_2 > 0$. Therefore, this SU has no intention to deviate from the individual optimal strategy defined in Definition~\ref{D:DuoIOS}.
\end{enumerate}
\section{Proof of Theorem \ref{T:DuoSuff}} \label{A:T_DuoSuff}
Since $\mathbf{q_d}(\mathbf{p}):[0, 1]^2\mapsto[0, 1]^2$ is differentiable, according to the contraction mapping\cite{Bertsekas1989}, the equilibrium  can be achieved and is stable for any initial $(p_1^{0}, p_2^{0})$ if the following condition is satisfied
\begin{align}
	\Gra \| \mathbf{q'_d}(\mathbf{p}) \| < 1, \quad  \forall \mathbf{p} \in [0, 1]^2, \label{E:normMax}
\end{align}
where $\mathbf{q'_d}(\mathbf{p})$ is the Jacobian matrix of $\mathbf{q_d}(\mathbf{p})$ and $\| . \|$ is a matrix norm. Using the $\infty$-norm, \eqref{E:normMax} is equivalent to
\begin{align}
 \Gra \bigl\vert  \FtheP \left( \bar{\Grq}(p_1) \right) \bigr\vert	< 1,	\quad	\forall p_1 \in \/[0, 1\/].	\label{E:DuoSuff1}
\end{align}
We have
\begin{align}
\bigl\vert \FtheP \left( \bar{\Grq}(p_1) \right) \bigr\vert = \Biggl\vert - \fthe \left( \bar{\Grq}(p_1) \right)  \bar{\Grq}(p_1)  \frac{\ExP{T( p_1 )}}{ \Ex{T(p_1)} -\Ex{X} }\Biggr\vert	&\leq \Biggl\vert \max_{\Grq \in \/[0,\theMax\/]}\fthe(\Grq) \Grq \Biggr\vert 
		      \Biggl\vert \max_{p_1 \in \/[0,1\/]} \frac{\ExP{T( p_1 )}}{\Ex{T(p_1)} -\Ex{X}}  \Biggr\vert	\nonumber\\
		&= \frac{\ExP{T( 1 )}}{\Ex{T(1)} -\Ex{X}}. \label{E:DuoSuff2}
\end{align}
From \eqref{E:DuoSuff1} and \eqref{E:DuoSuff2}, we complete the proof.

\section{Proof of Theorem \ref{T:DuoNash}} \label{A:T_DuoNash}
The revenue of \Oo is 
\begin{align}
	R_1(c_1, c_2) = c_1 p_1^* = c_1 \Fthe ( \bar{\Grq}(p_1^*) ) = \frac{c_1(c_2 - c_1)}{\theMax \bigl( \Ex{T(p_1^*)} - \Ex{X} \bigr) }.
\end{align}
Maximizing above with respect to $c_1$ by setting $\frac{\partial R_1}{\partial c_1} = 0$, we obtain the best response of \Oo
\begin{align}
	BR_1(c_2) = \frac{c_2}{2}.
\end{align}
Similarly, the revenue of \Oe is
\begin{align}
	R_2(c_1, c_2) = c_2 p_2^* = c_2 \Bigl( \Fthe \left( \Grq_2^* \right) - \Fthe \left( \bar{\Grq}(p_1^*) \right) \Bigr) = c_2 \left(  \frac{V - c_2}{\theMax \Ex{X}}  - \frac{c_2 - c_1}{\theMax \bigl( \Ex{T(p_1^*)} - \Ex{X} \bigr) } \right).
\end{align}
Maximizing above with respect to $c_2$ by setting $\frac{\partial R_2}{\partial c_2} = 0$, we obtain the best response of \Oe
\begin{align}
	BR_2(c_1) = \frac{V \bigl( \Ex{T(p_1^*)} - \Ex{X} \bigr) + c_1 \Ex{X}}{2 \ET}.
\end{align}
The Nash equilibrium strategy profile can be computed using the intersection of the best responses of both operators  as follows 
\begin{align}
	c_1^* &= BR_1(BR_2(c_1^*)) = \frac{ V \bigl( \Ex{T(p_1^*)} - \Ex{X} \bigr) }{4\ET - \Ex{X}},	\label{E:DuoPrice1}	\\
	c_2^* &= 2 c_1^*.	\label{E:DuoPrice2}
\end{align}
Substituting \eqref{E:DuoPrice1} and \eqref{E:DuoPrice2} into \eqref{E:DuoFixpoint}, we obtain
\begin{align}
	p_1^{\circledast} = \frac{ V}{\theMax \bigl( 4 \Ex{T(p_1^{\circledast})} - \Ex{X} \bigr)}.	\label{E:DuoFixPointII}
\end{align}
Using \eqref{E:qDelay}, the solution of the fixed-point equation \eqref{E:DuoFixPointII} can be found and is equal to \eqref{E:DuoPII}.


\bibliographystyle{IEEEtran}
\bibliography{IEEEabrv,mybib}

\begin{thebibliography}{10}
\providecommand{\url}[1]{#1}
\csname url@samestyle\endcsname
\providecommand{\newblock}{\relax}
\providecommand{\bibinfo}[2]{#2}
\providecommand{\BIBentrySTDinterwordspacing}{\spaceskip=0pt\relax}
\providecommand{\BIBentryALTinterwordstretchfactor}{4}
\providecommand{\BIBentryALTinterwordspacing}{\spaceskip=\fontdimen2\font plus
\BIBentryALTinterwordstretchfactor\fontdimen3\font minus
  \fontdimen4\font\relax}
\providecommand{\BIBforeignlanguage}[2]{{%
\expandafter\ifx\csname l@#1\endcsname\relax
\typeout{** WARNING: IEEEtran.bst: No hyphenation pattern has been}%
\typeout{** loaded for the language `#1'. Using the pattern for}%
\typeout{** the default language instead.}%
\else
\language=\csname l@#1\endcsname
\fi
#2}}
\providecommand{\BIBdecl}{\relax}
\BIBdecl

\bibitem{McHenry2005}
M.~A. McHenry, ``{NSF spectrum occupancy measurements project summary},''
  \emph{Shared Spectrum Company}, 2005.

\bibitem{Mitola1999}
J.~Mitola and G.~Maguire, ``Cognitive radio: making software radios more
  personal,'' \emph{Personal Communications, IEEE}, vol.~6, no.~4, pp. 13 --18,
  Aug. 1999.

\bibitem{Zhao2007}
Q.~Zhao and B.~Sadler, ``A survey of dynamic spectrum access,'' \emph{Signal
  Processing Magazine, IEEE}, vol.~24, no.~3, pp. 79 --89, May 2007.

\bibitem{Buddhikot2007}
M.~Buddhikot, ``Understanding dynamic spectrum access: Models, taxonomy and
  challenges,'' in \emph{Proc. IEEE DySPAN}, Dublin, Apr. 2007, pp. 649 --663.

\bibitem{Hossain2009}
E.~Hossain, D.~Niyato, and Z.~Han, \emph{Dynamic Spectrum Access and Management
  in Cognitive Radio Networks},\hskip 1em plus 0.5em minus 0.4em\relax New
  York, USA: Cambridge University Press, 2009.

\bibitem{Duan2011}
L.~Duan, J.~Huang, and B.~Shou, ``Investment and pricing with spectrum
  uncertainty: A cognitive operator's perspective,'' \emph{{IEEE} Trans. Mobile
  Comput.}, vol.~10, no.~11, pp. 1590--1604, Nov. 2011.

\bibitem{Kim2011a}
H.~Kim, J.~Choi, and K.~G. Shin, ``Wi-fi 2.0: Price and quality competitions of
  duopoly cognitive radio wireless service providers with time-varying spectrum
  availability,'' in \emph{Proc. IEEE INFOCOM}, Shanghai, Apr. 2011, pp.
  2453--2461.

\bibitem{Duan2012}
L.~Duan, J.~Huang, and B.~Shou, ``Duopoly competition in dynamic spectrum
  leasing and pricing,'' \emph{{IEEE} Trans. Mobile Comput.}, vol.~11, no.~11,
  pp. 1706--1719, Nov. 2012.

\bibitem{Jia2008a}
J.~Jia and Q.~Zhang, ``Competitions and dynamics of duopoly wireless service
  providers in dynamic spectrum market,'' in \emph{Proc. MOBIHOC 2008},\hskip
  1em plus 0.5em minus 0.4em\relax Hong Kong: ACM, 2008, pp. 313--322.

\bibitem{Niyato2009}
D.~Niyato, E.~Hossain, and Z.~Han, ``Dynamic spectrum access in ieee 802.22-
  based cognitive wireless networks: a game theoretic model for competitive
  spectrum bidding and pricing,'' \emph{Wireless Communications, IEEE},
  vol.~16, no.~2, pp. 16 --23, Apr. 2009.

\bibitem{Xing2007}
Y.~Xing, R.~Chandramouli, and C.~Cordeiro, ``Price dynamics in competitive
  agile spectrum access markets,'' \emph{Selected Areas in Communications, IEEE
  Journal on}, vol.~25, no.~3, pp. 613 -- 621, Apr. 2007.

\bibitem{Niyato2008}
D.~Niyato and E.~Hossain, ``Competitive pricing for spectrum sharing in
  cognitive radio networks: Dynamic game, inefficiency of nash equilibrium, and
  collusion,'' \emph{{IEEE} J. Sel. Areas Commun.}, vol.~26, no.~1, pp.
  192--202, Jan. 2008.

\bibitem{Yang2011}
L.~Yang, H.~Kim, J.~Zhang, M.~Chiang, and C.~W. Tan, ``Pricing-based spectrum
  access control in cognitive radio networks with random access,'' in
  \emph{Proc. IEEE INFOCOM}, Shanghai, Apr. 2011, pp. 2228 --2236.

\bibitem{Ileri2005}
O.~Ileri, D.~Samardzija, and N.~B. Mandayam, ``Demand responsive pricing and
  competitive spectrum allocation via a spectrum server,'' in \emph{Proc. IEEE
  DySPAN}, Baltimore, 2005, pp. 194--202.

\bibitem{Niyato2009a}
D.~Niyato, E.~Hossain, and Z.~Han, ``Dynamics of multiple-seller and
  multiple-buyer spectrum trading in cognitive radio networks: A game-theoretic
  modeling approach,'' \emph{Mobile Computing, IEEE Transactions on}, vol.~8,
  no.~8, pp. 1009 --1022, Aug. 2009.

\bibitem{Zhang2009}
J.~Zhang and Q.~Zhang, ``Stackelberg game for utility-based cooperative
  cognitive radio networks,'' in \emph{Proc. MobiHoc},\hskip 1em plus 0.5em
  minus 0.4em\relax New York: ACM, May 2009, pp. 23--32.

\bibitem{Naor1969}
P.~Naor, ``The regulation of queue size by levying tolls,''
  \emph{Econometrica}, vol.~37, no.~1, pp. 15--24, Jan. 1969.

\bibitem{Edelson1975}
N.~M. Edelson and D.~K. Hildebrand, ``Congestion tolls for poisson queuing
  processes,'' \emph{Econometrica}, vol.~43, no.~1, pp. 81--92, Jan. 1975.

\bibitem{Hassin2003}
R.~Hassin and M.~Haviv, \emph{{To Queue or Not to Queue: Equilibrium Behavior
  in Queueing Systems}},\hskip 1em plus 0.5em minus 0.4em\relax Springer, 2003.

\bibitem{Li2011}
H.~Li and Z.~Han, ``Socially optimal queuing control in cognitive radio
  networks subject to service interruptions: To queue or not to queue?''
  \emph{{IEEE} Trans. Wireless Commun.}, vol.~10, no.~5, pp. 1656--1666, May
  2011.

\bibitem{Economou2008}
A.~Economou and S.~Kanta, ``Equilibrium balking strategies in the observable
  single-server queue with breakdowns and repairs.'' \emph{Operations Research
  Letters}, vol.~36, no.~6, pp. 696--699, Jun. 2008.

\bibitem{Do2012}
C.~T. Do, N.~H. Tran, C.~S. Hong, and S.~Lee, ``Finding an individual optimal
  threshold of queue length in hybrid overlay/underlay spectrum access in
  cognitive radio networks.'' \emph{IEICE Transactions on Communications}, vol.
  95-B, no.~6, pp. 1978--1981, Jun. 2012.

\bibitem{Jagannathan2011}
K.~P. Jagannathan, I.~Menache, G.~Zussman, and E.~Modiano, ``Non-cooperative
  spectrum access: the dedicated vs. free spectrum choice,'' in
  \emph{Proc. MOBIHOC 2011},\hskip 1em plus 0.5em minus 0.4em\relax Paris, ACM, 2011, pp.
  10:1--10:11.

\bibitem{Manshaei2008}
M.~H. Manshaei, J.~Freudiger, M.~Felegyhazi, P.~Marbach, and J.-P. Hubaux, ``On
  wireless social community networks,'' in \emph{Proc. IEEE INFOCOM}, Pheonix,
  AZ, Apr. 2008, pp. 1552--1560.

\bibitem{Chau2010}
C.-K. Chau, Q.~Wang, and D.-M. Chiu, ``On the viability of paris metro pricing
  for communication and service networks,'' in \emph{Proc. IEEE INFOCOM}, San
  Diego, CA, Mar. 2010, pp. 929--937.

\bibitem{Ren2011}
S.~Ren, J.~Park, and M.~van~der Schaar, ``User subscription dynamics and
  revenue maximization in communications markets,'' in \emph{Proc. IEEE
  INFOCOM}, Shanghai, Apr. 2011, pp. 2696 --2704.

\bibitem{Gibbens2000}
R.~Gibbens, R.~Mason, and R.~Steinberg, ``Internet service classes under
  competition,'' \emph{{IEEE} J. Sel. Areas Commun.}, vol.~18, no.~12, pp.
  2490--2498, Dec. 2000.

\bibitem{Geirhofer2008}
S.~Geirhofer, L.~Tong, and B.~M. Sadler, ``Cognitive medium access:
  Constraining interference based on experimental models,'' \emph{{IEEE} J.
  Sel. Areas Commun.}, vol.~26, no.~1, pp. 95--105, Jan. 2008.

\bibitem{Shiang2008}
H.-P. Shiang and M.~van~der Schaar, ``Queuing-based dynamic channel selection
  for heterogeneous multimedia applications over cognitive radio networks,''
  \emph{{IEEE} Trans. Multimedia}, vol.~10, no.~5, pp. 896--909, Aug. 2008.

\bibitem{Li2008}
X.~Li and S.~Zekavat, ``Traffic pattern prediction and performance
  investigation for cognitive radio systems,'' in \emph{Proc. IEEE WCNC},  Las Vegas, Apr.
  2008, pp. 894 --899.

\bibitem{Stevenson2009}
C.~Stevenson, G.~Chouinard, Z.~Lei, W.~Hu, S.~Shellhammer, and W.~Caldwell,
  ``{IEEE 802.22: The first cognitive radio wireless regional area network
  standard},'' \emph{IEEE Communications Magazine}, vol.~47, no.~1, pp. 130
  --138, Jan. 2009.

\bibitem{Bertsekas1992}
D.~Bertsekas and R.~Gallager, \emph{Data networks (2nd ed)},\hskip 1em plus
  0.5em minus 0.4em\relax NJ, USA: Prentice-Hall, Inc.,
  1992.

\bibitem{Cox1967}
D.~R. Cox, \emph{Renewal Theory},\hskip 1em plus 0.5em minus 0.4em\relax Butler
  \& Tanner Ltd, London, 1967.

\bibitem{Evans2001}
G.~Evans and S.~Honkapohja, \emph{Learning and expectations in
  macroeconomics},\hskip 1em plus 0.5em minus 0.4em\relax Princeton Univ.
  Press, 2001.

\bibitem{Osborne1994}
M.~J. Osborne and A.~Rubinstein, \emph{{A Course in Game Theory}},\hskip 1em plus 0.5em minus 0.4em\relax The MIT Press, Jul. 1994.

\bibitem{Han2011}
Z.~Han, D.~Niyato, W.~Saad, T.~Basar, and A.~Hjorungnes, \emph{Game Theory in
  Wireless and Communication Networks: Theory, Models, and Applications},\hskip
  1em plus 0.5em minus 0.4em\relax New York, NY, USA: Cambridge University
  Press, 2011.

\bibitem{Bertsekas1989}
D.~P. Bertsekas and J.~N. Tsitsiklis, \emph{Parallel and Distributed
  Computation: Numerical Methods},\hskip 1em plus 0.5em minus 0.4em\relax Englewood Cliffs, NJ:
  Prentice-Hall, 1989.

\end{thebibliography}


\end{document}